\documentclass[12pt]{article}
\usepackage{epsfig}
\usepackage[numbers,sort&compress]{natbib}
\usepackage{amsmath,amssymb}
\usepackage{float}
\usepackage[english]{babel}
\usepackage{graphicx}

\def\Title#1{\begin{center} {\Large {\bf #1} } \end{center}}

\begin{document}

\Title{Recent CMS Results on Diffraction}

\bigskip\bigskip

\begin{raggedright}  

\vspace*{-0.4cm}
{\it Roland Beno\^it\index{Roland, B.} on behalf of the CMS Collaboration} \\
{\it DESY - CMS \\
Notkestrasse 85, 22603 Hamburg \\
Germany
}
\bigskip\bigskip

\end{raggedright}

\vspace*{-1.2cm}
\begin{abstract}
  
Recent CMS results on diffraction are presented. These include the measurements of the soft diffractive cross sections, of the forward rapidity gap cross section,
of the diffractive dijet cross section, the measurement of a large rapidity gap in $W$ and $Z$ boson events and the measurement of the pseudorapidity distribution of charged particles
in a single diffractive enhanced sample. This last measurement is the first common result of the CMS and TOTEM collaborations. Some prospects of common CMS-TOTEM data taking are also discussed.
\end{abstract}


\vspace*{-0.8cm}

\section{Diffraction}

Diffractive reactions represent a sizable fraction of the total inelastic cross section $\sigma_{tot}$ in hadron-hadron scattering,
of the order of $25\%$ of $\sigma_{tot}$ \cite{Cartiglia:2013jya}. These reactions are characterized by a color neutral 
$t$-channel exchange carrying the quantum numbers of the vacuum, where $t$ is the four-momentum squared exchanged 
in the collision \cite{Barone:2002cv}. In proton-proton ($p p$) collisions, single diffractive (SD) dissociation $pp \rightarrow p X$ corresponds to the process where one of the protons emerges
intact in the final state, while the other is scattered into a low-mass system $X$. In double diffractive (DD) dissociation $pp \rightarrow XY$, 
both of the protons are scattered into low-mass states $X$ and $Y$, whereas in central diffractive (CD) dissociation, $pp \rightarrow pXp$, both
protons emerge intact from the collision, with a low-mass system $X$ produced centrally. In all cases, the energy of the 
outgoing protons or states $X,Y$ is approximately equal to that of the incoming protons, to within a few percent.
Diffractive events are characterized by the presence of at least one large rapidity region $\Delta y$ devoid of hadronic activity \cite{Arneodo:2005kd}, 
the so-called large rapidity gap (LRG), where the rapidity $y$ is defined as $y = \mbox{ln}[(E+p_Z)/(E-p_Z)]$, $E$ and $p_Z$ being the energy
and longitudinal momentum of the final state particle, respectively. The pseudorapidity $\eta = - \mbox{ln} [ \mbox{tg} (\theta/2)]$, with
$\theta$ the polar angle of the particle, is often used experimentally instead of rapidity. Both variables are equal in the limit of a 
massless particle. \newline
  
Diffractive hadron-hadron collisions can be described in the framework of Regge theory~\cite{Barone:2002cv,Collins:1977jy}. In this framework, the 
$t$-channel process is governed by the exchange of resonances sitting on a so-called Regge trajectory $\alpha(t) = \alpha + \alpha' t$.  
A resonance of mass $m$ sitting on the trajectory at $t = m^2$ has an angular momentum of value $\alpha(m^2)$, and a Regge trajectory therefore interpolates 
between resonances of increasing angular momentum. The contribution of a trajectory to the total inelastic cross section is given by $\sigma_{tot}(s) \propto s^{\alpha-1}$, 
where $s$ is the centre-of-mass energy squared of the collision and $\alpha$ the intercept of the trajectory. In the Regge framework, diffraction is characterized by the exchange 
of a specific trajectory, the so-called Pomeron trajectory, which has the quantum numbers of the vacuum and an intercept $\alpha$ larger than one. \newline

Hard diffractive electron-proton ($e p$) collisions can be described by perturbative Quantum Chromodynamics (QCD) in the framework of collinear factorization
\cite{Collins:1997sr,Berera:1995fj}. In this framework, the cross section of a hard diffractive process is given by the
convolution of a parton-level cross section, which is process-dependent, with the diffractive parton distribution
functions (dPDFs), which are process-independent. The evolution of the dPDFs with the four-momentum squared exchanged at the proton vertex, $Q^2$,
is governed by the DGLAP evolution equations \cite{Gribov:1972ri,Lipatov:1974qm,Altarelli:1977zs,Dokshitzer:1977sg}.
In the parton model, these functions can be interpreted at leading-order as conditional probabilities to find a parton
in the proton, carrying a fraction $x$ of its longitudinal momentum, at a given value of $Q^2$, under the assumption that
a fast leading proton is present in the final state. Hard diffractive $e p$ collisions have been studied extensively
at the $e p$ HERA collider through the measurement of both inclusive and exclusive processes
\cite{Aktas:2006hy,Chekanov:2009aa,Chekanov:2008fh,Aaron:2010su,Aaron:2012hua,Aaron:2012ad,Aaron:2011mp}. The dPDFs have been determined by the HERA experiments
\cite{Aktas:2006hy,Chekanov:2009aa} by means of QCD fits to inclusive diffractive deep inelastic scattering data. \newline

Hard diffractive processes such as the diffractive production of jets \cite{Affolder:2000vb,Abbott:1999km,Aaltonen:2012tha}, or $W$ and $Z$ bosons
 \cite{Aaltonen:2010qe} have been studied at hadron colliders. The measurement of diffractive dijet production 
in $p\bar{p}$ collisions from the CDF collaboration at the Tevatron \cite{Affolder:2000vb} was the first to show an important disagreement 
with the theoretical expectations based on the dPDFs determined from HERA data.  
The fraction of diffractive dijet events is a factor 3 to 10 smaller than the expectation,
with the reduction becoming more important with increasing value of the struck parton momentum fraction. Collinear factorization
can not be applied to diffractive hadron-hadron collisions \cite{Kaidalov:2009fp,Klasen:2010we}, because additional soft or semi-hard 
interactions between the spectator partons, so-called multiple parton interactions (MPI), can occur and produce hadronic activity 
that lowers the probability for the rapidity gap to form. 
With this picture in mind, the reduction factor is often referred to as the rapidity gap survival probability~$S^2$.          
Various models have been investigated to determine the expected behaviour of $S^2$ at the LHC centre-of-mass energy $\sqrt{s} = 7$~TeV.
These models mainly differ on the type of the Pomeron amplitude 
describing the soft re-scatterings at the origin of the rapidity gap suppression \cite{Gotsman:2005rt,Khoze:2006gg,Gotsman:2008mm,Gotsman:2011xc}. 
Most of the theoretical calculations predict a value of $S^2 \simeq 0.05$ for hard diffractive processes at LHC energies, 
but values as low as 0.004 and as high as 0.23 have been proposed \cite{Bartalini:2010su}. 
\newline

Diffractive factorization breaking is intimately related to multiple scatterings in hadron-hadron collisions. 
If one assumes that re-scatterings are independent of each other, 
the distribution of the number of re-scatterings at an impact parameter $b$ can be modelled by the Poisson statistics.
Within this simple model, the gap survival probability is directly related to the mean number of re-scatterings at an impact parameter $b$, $\mu(b)$,
through the relation \cite{Alekhin:2005dy}: 
\begin{equation}
S^2 = \frac{\int db \, \frac{d\sigma(b)}{db} \, e^{-\mu(b)}}{\int db \, \frac{d\sigma(b)}{db}},
\end{equation}
where $\sigma(b)$ is the cross section for the hard diffractive process as a function of $b$.

\section{Diffraction in CMS}

The study of diffraction at the LHC is of interest in several aspects. As diffractive reactions represent a sizeable fraction of the total proton-proton cross section,
their contribution to the Underlying Event (UE), which denotes all the particles produced in a hard $pp$ scattering apart from the hard scattering system itself, is substantial.
The CMS potential to achieve precise Standard Model measurements and to search for new physics is affected by the Underlying Event activity. Soft diffractive interactions, 
in which no hard scale is present, represent the bulk of the diffractive events. Given the importance of its contribution to the UE and the need to constrain its phenomenological description, 
the measurement of the soft diffractive cross section is therefore of primary importance. \newline

The measurement of hard diffractive processes, in which the transverse momentum or the mass of the produced system defines the hard scale of the event, provides an important test of QCD 
and probes the low-$x$ structure of the proton, with $x$ the fractional momentum of the struck parton with respect to that of the incoming proton. In the low-$x$ region, the strong rise of 
the gluon density is expected to be tamed by recombination processes, leading to the saturation of the gluon density. The observation of this non-linear effect, which is not included in the 
DGLAP evolution equations \cite{Gribov:1972ri,Lipatov:1974qm,Altarelli:1977zs,Dokshitzer:1977sg}, would shed light on the asymptotic high-energy behaviour of QCD. \newline

Finally, the measurement of the rapidity gap survival probability, which quantifies the suppression of the diffractive cross section in $pp$ collisions with respect to $ep$ scattering, is also of
primary importance. The survival probability is poorly constrained, 
from both an experimental and a theoretical point of view, and its precise measurement would enable us to better understand the soft multiple parton interactions  
that occur in high-energy $pp$ collisions. MPI could fake or deteriorate signals for new physics, and a better knowledge of their dynamics would shed light on soft QCD, and would benefit precise 
measurements and searches for new physics as well. \newline

During the LHC Run 1, diffractive processes have been mainly selected by requiring a large rapidity gap in the event, and all the CMS results presented in the following sections make use 
of the rapidity gap tagging to select diffractive events. The successful collaboration between the CMS and TOTEM experiments has recently enabled to complement the central CMS measurement 
with the scattered proton information measured with the TOTEM Roman pots in dedicated common runs during 2012 and 2013. Tagging the scattered proton(s) enables us to measure the four-momentum 
squared exchanged at the proton vertex, $t$, as well as the fractional momentum loss of the proton, $\xi$, when studying single diffractive dissociation $pp \to p X$ and central exclusive production 
(CEP) $pp \to p X p$. The study of these two processes with the CMS central detector only is affected by an irreducible proton dissociation background, which arises from events in which the scattered 
proton dissociates into a low-mass system $Y$ that escapes undetected. Tagging the scattered proton(s) makes it possible to significantly reduce the proton dissociation background 
when studying single diffractive dissociation and central exclusive production. In the case of CEP, tagging the scattered protons allows to reconstruct the mass of the centrally produced system 
with a resolution not achievable with the CMS detector only. The current ongoing analyses that make use of the proton tagging are dedicated to the study of single diffractive production
of a dijet system, $J/\Psi$ or $D^0$ meson. While these analyses are using the information from both CMS and TOTEM, they suffer from low statistics and from the absence of dedicated triggers making
use of the CMS central detector and TOTEM information. \newline

Sections \ref{sec:detectorCMS} and \ref{sec:detectorTOTEM} give a short description of the CMS and TOTEM detectors, emphasizing on the subdetectors used in the diffractive analyses.
The CMS results are presented in Sections \ref{sec:soft} to \ref{sec:eta}. These include the measurements of the soft diffractive cross sections~\cite{CMS:2013mda}, of the forward rapidity gap 
cross section~\cite{CMS:2013mda}, of the diffractive dijet cross section \cite{Chatrchyan:2012vc}, the measurement of a large rapidity gap in $W$ and $Z$ boson events \cite{Chatrchyan:2011wb} 
and the measurement of the pseudorapidity distribution of charged particles in a single diffractive enhanced sample~\cite{Chatrchyan:2014qka}. This last measurement is the first common result 
of the CMS and TOTEM collaborations. Some prospects of common CMS-TOTEM data taking are discussed in Sec.\ref{sec:prospects}, and a summary is given in Sec.\ref{sec:summary}.

\section{The CMS detector}
\label{sec:detectorCMS}

A detailed description of the CMS detector can be found in Ref.~\cite{Chatrchyan:2008aa}. The central feature of the apparatus is a superconducting solenoid of 6 m internal diameter, 
providing a 3.8 T axial field. Within the field volume are a silicon pixel and strip tracker, a crystal electromagnetic calorimeter (ECAL) and a brass/scintillator hadron calorimeter (HCAL). 
Muons are measured in gas-ionization detectors embedded in the steel flux return yoke. In addition to the barrel and endcap detectors, CMS has extensive forward calorimetry. \newline

The CMS experiment uses a right-handed coordinate system, with the origin at the nominal interaction point (IP), the $x$-axis pointing to the center of the LHC, the $y$-axis pointing up 
(perpendicular to the plane of the LHC ring), and the $z$-axis along the anticlockwise-beam direction. The polar angle $\theta$ is measured from the positive $z$-axis and 
the azimuthal angle $\phi$  is measured in the $x$--$y$ plane. \newline

The tracker measures charged particles within the pseudorapidity range $|\eta| < 2.5$. It provides an impact parameter resolution of $\sim 15 \, \mu$m and a transverse momentum resolution 
of about $1.5\%$ for $100 \, \mbox{GeV}/c$ particles. ECAL and HCAL provide coverage in pseudorapidity up to $|\eta| < 3$ in the barrel region and two endcap regions. The ECAL has an energy resolution 
of better than 0.5$\%$ above 100 GeV. The HCAL, when combined with the ECAL, measures jets with an energy resolution $\Delta E / E \approx 100 \% / \sqrt{E \mbox{(GeV)}} \oplus 5\% $.
The calorimeter cells are grouped in projective towers, of granularity $\Delta \eta \times \Delta \phi = 0.087 \times 0.087$ at central rapidities and $0.175 \times 0.175$ at forward rapidities.
The hadronic forward (HF) calorimeters cover the pseudorapidity region $2.9 < |\eta| < 5.2$. They consist of steel absorbers and embedded radiation-hard quartz fibres, which provide a fast collection 
of the Cherenkov light. Calorimeter cells are formed by grouping bundles of fibres. Clusters of these cells form a calorimeter tower. There are 13 towers in $|\eta|$, each with a size
$\Delta \eta \approx 0.175$, except for the lowest- and highest-$|\eta|$ towers with $\Delta \eta \approx 0.1$ and $\Delta \eta \approx 0.3$, respectively. The azimuthal segmentation
$\Delta \phi$ of all towers is $10^\circ$, except for the ones at highest-$|\eta|$, which have $\Delta \phi = 20^\circ$. 
More forward angles, $-6.6 < \eta < -5.2$, are covered by the CASTOR calorimeter \cite{Andreev:2010zzb}, which is located only on the negative $z$-side of CMS, at 14.37 m from the IP. 
The calorimeter is segmented in 16 $\phi$-sectors and 14 $z$-modules, corresponding to a total of 224 cells. Each cell consists of quartz plates embedded in tungsten absorbers, 
with $45^\circ$ inclination with respect to the beam axis. Air core light guides provide a fast collection of the Cherenkov light. The first two modules, which have an absorber thickness half of
that of the other modules, are used to detect electromagnetic showers. The full calorimeter has a depth of 10.5 nuclear interaction lengths.

\section{The TOTEM detector}
\label{sec:detectorTOTEM}

The TOTEM experiment \cite{Anelli:2008zza,Antchev:2013hya} is composed of three subdetectors: the Roman pots, and the T1 and T2 telescopes. The two T2 telescopes are placed symmetrically on each side 
of the IP at about $|z| = 14$ m. They detect charged particles produced in the pseudorapidity region $5.3 < |\eta| < 6.5$, with full azimuthal acceptance. Each telescope consists of two half-arms, 
with each half-arm composed of 10 semicircular planes of triple-GEM (gas electron multiplier) chambers, arranged within a 40 cm space along the $z$-axis. Every chamber has a double-layered readout 
board containing two columns of 256 concentric strips to measure the radial coordinate and a matrix of 1560 pads, each covering  $\Delta \eta \times \Delta \phi = 0.06 \times 0.018$, to measure 
the azimuthal coordinate. The radial and azimuthal coordinate resolutions are about $110 \, \mu$m and $1^\circ$, respectively \cite{Bagliesi:2010zz}. Local angles of single tracks are reconstructed 
with an average resolution of 0.5 mrad, and the track pseudorapidity resolution for charged particles is better than 0.05 \cite{Aspell:2012ux}, once the track is identified as coming from the vertex.

\section{Soft diffractive cross sections}
\label{sec:soft}

\begin{figure}[htb]
\centering
\vspace*{-0.2cm}
\includegraphics[scale=0.35]{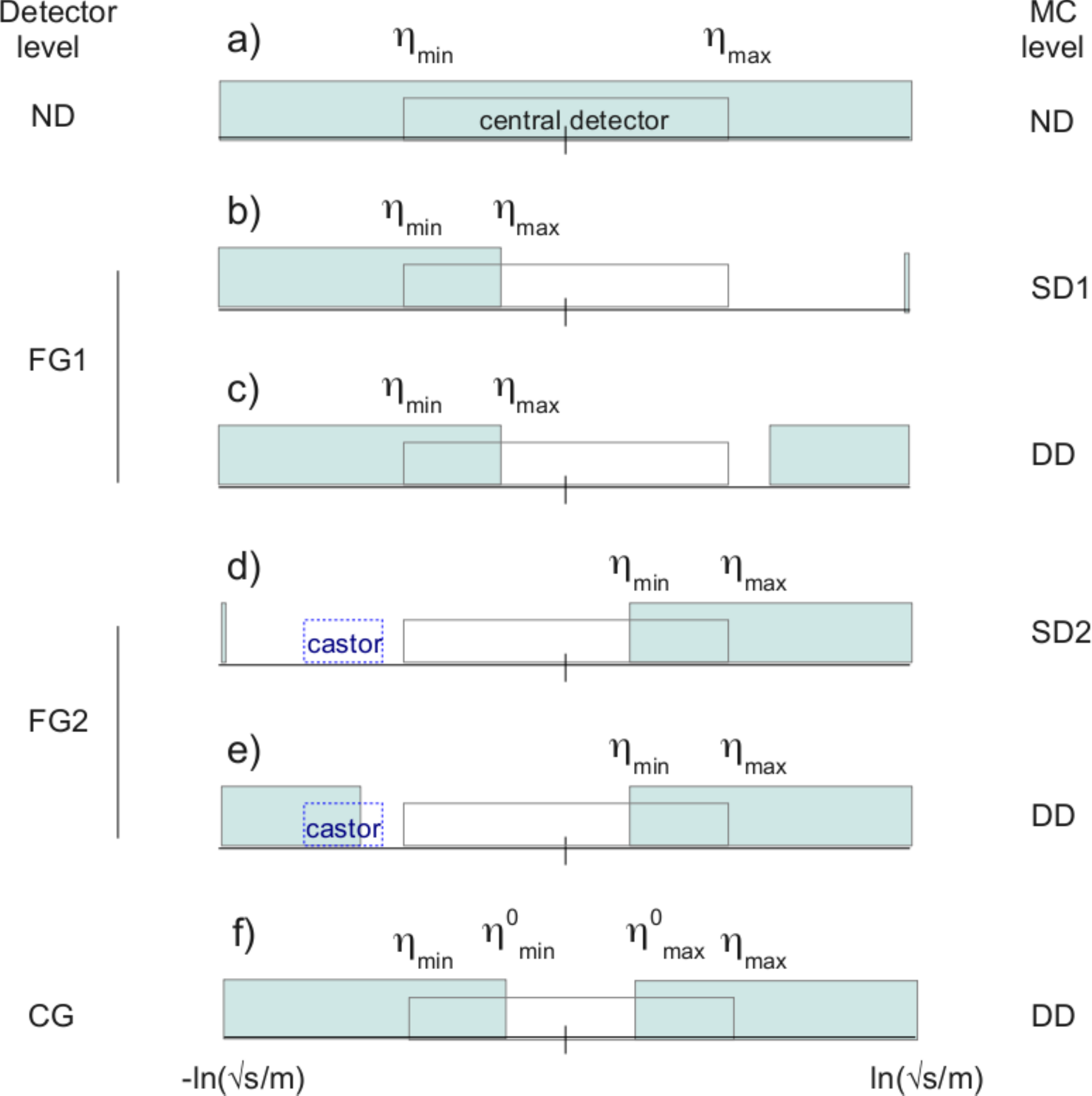}
\caption{Diffractive topologies defined experimentally. Detector level: non diffractive events (ND), diffractive events with a forward pseudorapidity gap on the positive (FG1) or negative 
(FG2) $\eta$-side of the detector, or with a central gap (CG). MC level: ND events $pp\rightarrow X$, SD1 (SD2) events $pp\rightarrow Xp$ ($pp\rightarrow pX$) and DD events $pp\rightarrow XY$.}
\label{fig:exp-definition}
\end{figure}

The measurement of the SD and DD dissociation cross sections~\cite{CMS:2013mda} is based on a sample of $pp$ collisions at $\sqrt{s} = 7$ TeV selected online by a minimum-bias (MB) trigger.
The data sample was collected when the LHC was operating in a low pileup scenario and corresponds to an integrated luminosity of 16.2 $\mu\mbox{b}^{-1}$ with an average number of inelastic 
$pp$ collisions per bunch crossing of 0.14. This minimum-bias sample is dominated by nondiffractive (ND) events in which the production of the final state particles extends over the entire
available phase-space. The presence of a rapidity gap in a ND event is due to statistical fluctuations in the production of the final state particles, and large rapidity gaps in ND events 
are therefore exponentially suppressed. In contrast, diffractive events are characterized by the presence of at least one non-exponentially suppressed LRG. The diffractive 
topologies defined experimentally according to the position of the LRG are presented in figure~\ref{fig:exp-definition}, where the open boxes represent the central CMS detector ($|\eta|\lesssim 4.7$), 
the dotted open boxes the CASTOR calorimeter and the full boxes the final state systems. \\

\begin{figure}[htb]
\centering
\vspace*{-0.3cm}
\hspace*{-0.8cm}
\includegraphics[scale=0.9]{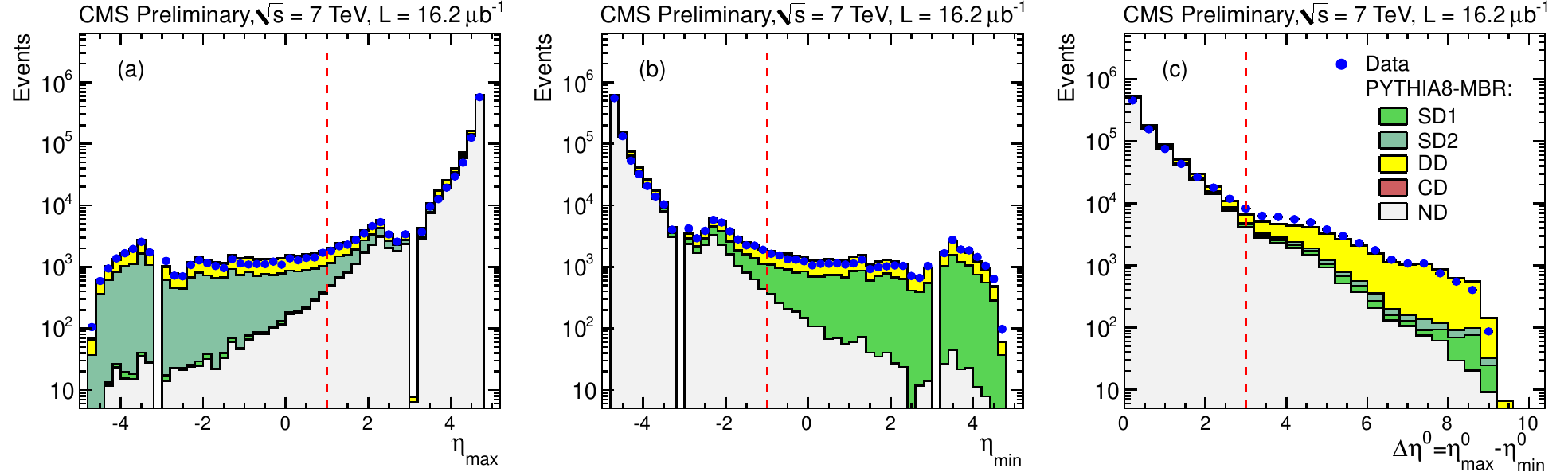}
\vspace*{-0.5cm}
\caption{Detector-level distributions of $\eta_{max}$ (a), $\eta_{min}$ (b) and $\Delta \eta^0$ (c) for the minimum-bias sample. The data are compared to the predictions of the MC event generator 
{\sc PYTHIA}8-MBR. Contributions are shown for each process separately. The dashed vertical lines represent the selections $\eta_{max} < 1$, $\eta_{min} > -1$ and $\Delta \eta^0 > 3$, respectively.}
\label{fig:eta-det}
\end{figure}

The samples FG1 and FG2 (Forward Gap) correspond to diffractive events with a forward pseudorapidity gap reconstructed at the edge of the detector, on the positive (FG1) or negative (FG2) $\eta$-side. 
The CG sample (Central Gap) corresponds to diffractive events with a rapidity gap reconstructed in the detector around $\eta = 0$. For the experimental topology FG1 (FG2), the forward rapidity gap 
is related to the variable $\eta_{max}$ $(\eta_{min})$ defined as the highest (lowest) $\eta$ of the particle candidate reconstructed in the central detector. For the CG topology, the pseudorapidity gap 
is expressed as $\Delta \eta^0 = \eta^0_{max} - \eta^0_{min}$, with $\eta^0_{max}$ $(\eta^0_{min})$ the closest-to-zero $\eta$ value of the particle candidate reconstructed on the positive (negative) 
$\eta$-side of the detector. The detector-level distributions of $\eta_{max}$, $\eta_{min}$ and $\Delta \eta^0$ are presented in figure~\ref{fig:eta-det} and compared to the predictions of the Monte Carlo 
(MC) event generator {\sc PYTHIA}8-MBR~\cite{Ciesielski:2012mc}. The ND contribution dominates the distributions at low values of the gap size, while the flattening of the exponential behaviour with 
increasing value of the gap is due to the contribution of the diffractive events. These dominate the distributions at low values of $\eta_{max}$, high values of $\eta_{min}$ 
or high values of $\Delta \eta^0$.

 \newpage

The diffractive-enhanced samples FG1 and FG2 are defined by the selections $\eta_{max} < 1$ and $\eta_{min} > -1$, respectively. This condition corresponds to the presence of a forward LRG 
of at least 3.7 units in pseudorapidity. The diffractive-enhanced CG sample is defined by the selection $\Delta \eta^0 > 3$.
This sample, for which both diffractive masses are required to be in the central detector, is dominated by ND and DD events. The diffractive-enhanced samples FG1 ($\eta_{max} < 1$) 
and FG2 ($\eta_{min} > -1$) receive contributions from SD and DD processes in approximately equal amount. The DD contribution originates from events in which one of the dissociated systems has a low-mass 
and is produced outside of the acceptance of the central detector ($|\eta|\lesssim 4.7$), as illustrated in the figure~\ref{fig:exp-definition} c) and e). In the case of the FG1 sample, no detector is 
present on the positive $\eta$-side of CMS to measure a low-mass dissociated system that escapes detection in the central detector, and the FG1 sample is regarded as a control sample. For the FG2 sample, 
the CASTOR calorimeter can be used to tag the presence of a low-mass dissociated system in the pseudorapidity range $-6.6 < \eta < -5.2$, to further divide the FG2 sample into a SD-enhanced subsample 
and a DD-enhanced subsample, according to the absence or the presence of activity in CASTOR, respectively. \newline

\begin{figure}[htb]
\centering
\hspace*{-0.8cm}
\includegraphics[scale=0.9]{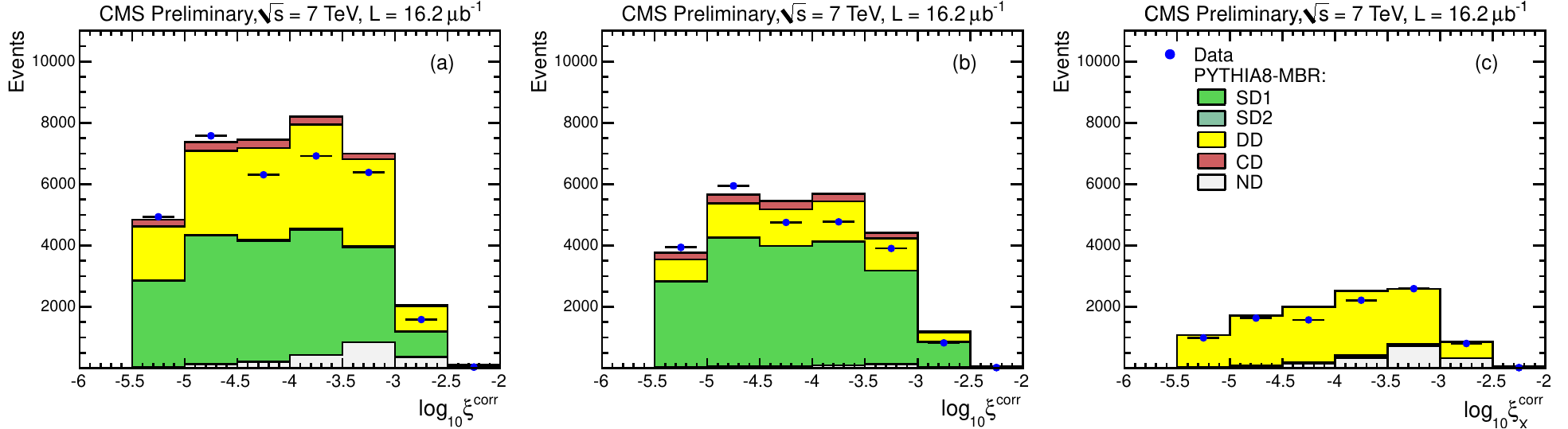}

\vspace*{0.1cm}
\hspace*{-0.8cm}
\includegraphics[scale=0.9]{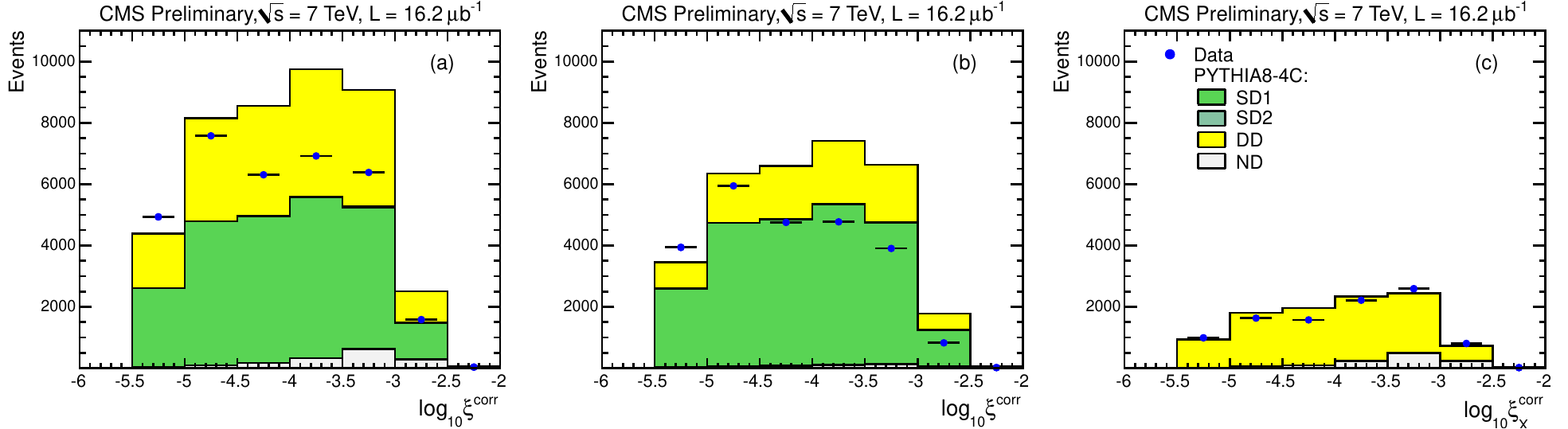}
\vspace*{-0.2cm}
\caption{Detector-level distributions of the variable $\xi$ for the entire FG2 sample (a), and for the subsamples without activity in CASTOR (b) and with activity in CASTOR (c).
The data are compared to predictions of the MC event generators {\sc PYTHIA}8-MBR (top) and {\sc PYTHIA}8-4C (bottom). Contributions are shown for each process separately.}
\label{fig:logxi-det}
\end{figure}

The detector-level distributions of $\mbox{log}_{10} \xi$ are presented in figure~\ref{fig:logxi-det} for the entire FG2 sample (a), and for the subsamples without activity in CASTOR (b) and with activity 
in CASTOR (c). The FG2 sample is clearly separated into a SD-enhanced subsample (b) and a DD-enhanced subsample (c). The data are compared to predictions of the MC event generators {\sc PYTHIA}8-MBR 
and {\sc PYTHIA}8~\cite{Sjostrand:2007gs} with tune 4C~\cite{Corke:2010yf}. The fractional momentum loss of the proton, $\xi$, is reconstructed from the energy $E^i$ and the longitudinal momentum $p_z^i$ 
of the particle candidates measured in the central detector, according to the relation $\xi = \sum(E^i-p_z^i)/\sqrt{s}$. The MC event generator {\sc PYTHIA}8-MBR is found to give a better description 
of the data and is used to determine the diffractive cross sections.  \\

\begin{figure}[h!]
\centering
\includegraphics[scale=0.8]{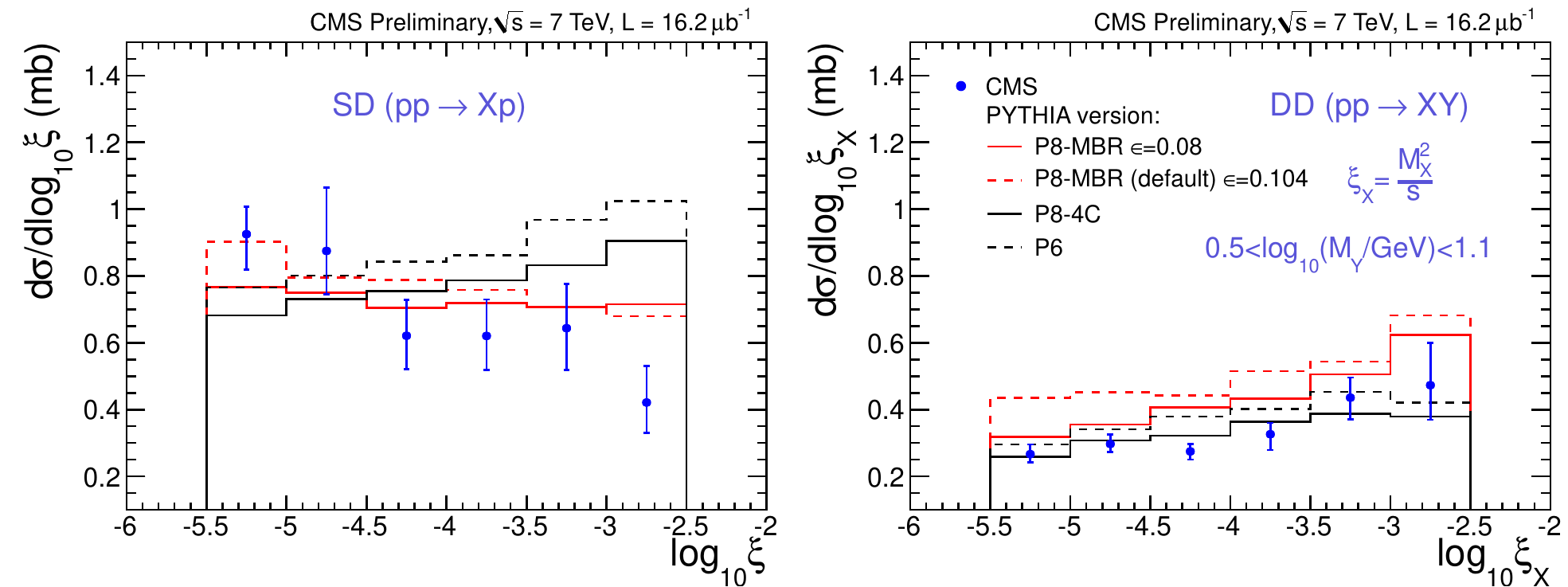}
\vspace*{-0.2cm}
\caption{The SD and DD differential cross sections as a function of $\xi$. The measurements are compared to predictions of the MC event generators {\sc PYTHIA}8-MBR, {\sc PYTHIA}8-4C 
and {\sc PYTHIA}6-D6T. The error bars represent the statistical and systematic uncertainties added in quadrature.}
\label{fig:SDDD}
\end{figure}

The SD and DD differential cross sections as a function of $\mbox{log}_{10} \xi$ are determined from the subsamples without activity in CASTOR and with activity in CASTOR, respectively.
The correction for the detector acceptance and the subtraction of the background contributions are determined with the {\sc PYTHIA}8-MBR MC event generator. The dominant background contribution 
to the SD cross section originates from DD events. For the DD cross section, the dominant background contribution originates from ND events.
The cross sections are measured in the range $-5.5 < \mbox{log}_{10} \xi < -2.5$. For the DD cross section, the mass of the system $Y$ is limited to the region 
$0.5 < \mbox{log}_{10}(M_Y/\mbox{GeV}) < 1.1$ corresponding to the acceptance of the CASTOR calorimeter. The dominant sources of systematic uncertainties are the energy scale of the HF calorimeter  
and the modelling of hadronization and diffraction. The measured cross sections are presented in figure~\ref{fig:SDDD} and compared to predictions of the MC event generators {\sc PYTHIA}8-MBR, 
{\sc PYTHIA}8-4C and {\sc PYTHIA}6~\cite{Sjostrand:2006za} with tune D6T~\cite{Field:2008zz}. The predictions from {\sc PYTHIA}8-MBR are shown for two values of the Pomeron intercept, $\alpha = 1.08$ 
and $\alpha = 1.104$. Both values describe the SD cross section within uncertainties, while the DD cross section is better described by the smaller value of the intercept. The predictions 
from {\sc PYTHIA}8-4C and {\sc PYTHIA}6-D6T, based on the Schuler-Sjostrand model, give a good description of the DD cross section but fail to describe the falling behavior of the SD cross section.\newpage

\vspace*{0.5cm}

 \newpage

\begin{figure}[ht!]
\centering
\includegraphics[scale=0.4]{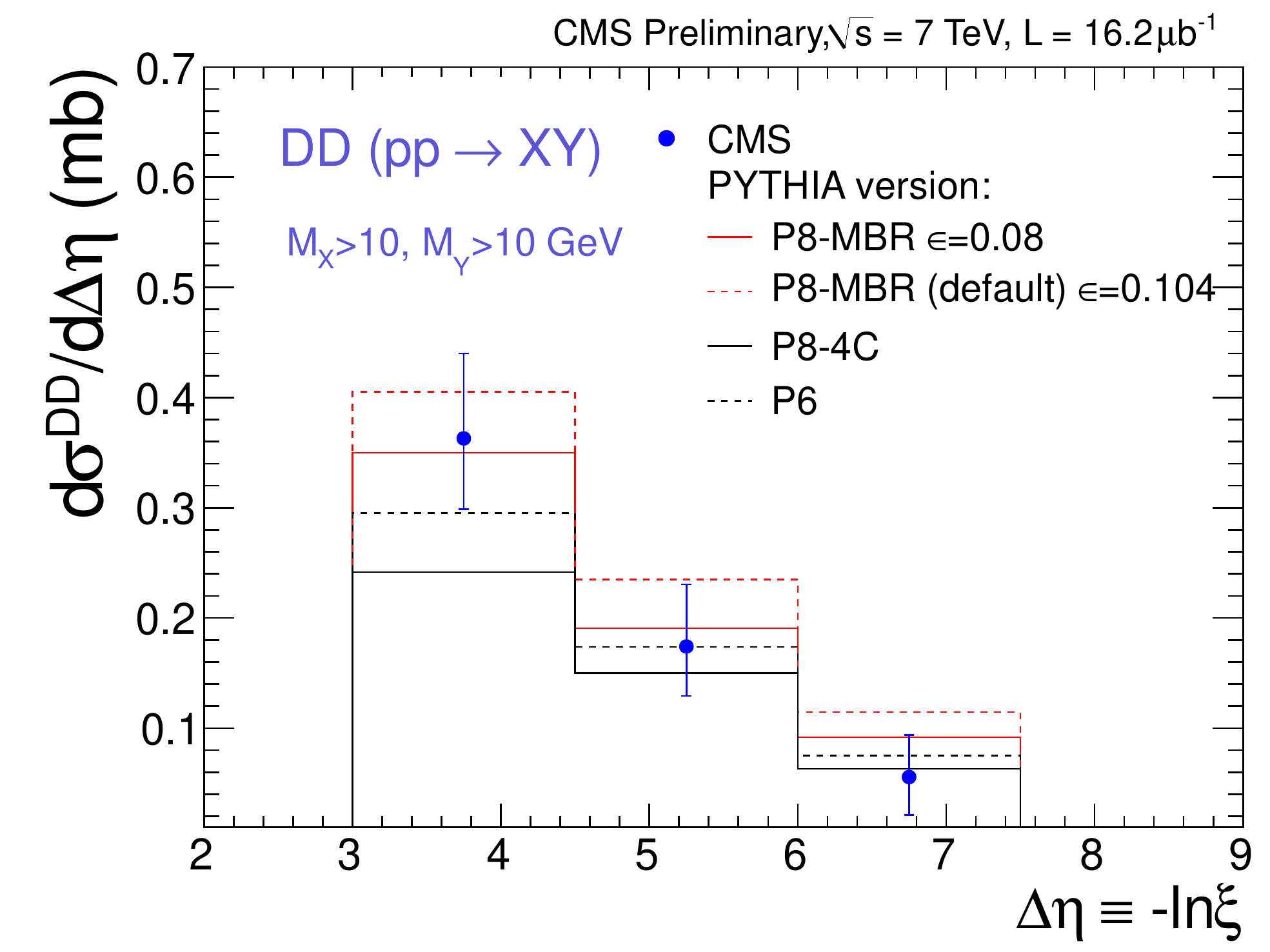}
\vspace*{-0.3cm}
\caption{The DD differential cross section as a function of $\Delta \eta$. The measurement is compared to predictions of the MC event generators {\sc PYTHIA}8-MBR, {\sc PYTHIA}8-4C 
and {\sc PYTHIA}6-D6T. The error bars represent the statistical and systematic uncertainties added in quadrature.}
\label{fig:DD}
\end{figure}

The DD differential cross section as a function of the size of the pseudorapidity gap, $\Delta \eta$, is determined from the CG sample. The variable $\Delta \eta$ is defined 
as $\Delta \eta = -\mbox{log}\, \xi$, with $\xi~=~M^2_X~\cdot~M^2_Y~/~(s~\cdot m^2_p)$, $M_X$ ($M_Y$) the invariant mass of the system $X$ ($Y$) and $m_p$ the proton mass. 
The cross section is measured in the range $\Delta \eta > 3$, $M_X > 10$ GeV and $M_Y > 10$~GeV. The correction for the detector acceptance and the subtraction of the background are determined 
with the {\sc PYTHIA}8-MBR MC event generator. The dominant background contribution originates from ND events. The measured cross section is presented in figure~\ref{fig:DD} and compared to 
predictions of the MC event generators {\sc PYTHIA}8-MBR, {\sc PYTHIA}8-4C and {\sc PYTHIA}6-D6T. The predictions give a reasonable description of the data within uncertainties.

The total SD cross section at $\sqrt{s} = 7$ TeV integrated over the region $-5.5 < \mbox{log}_{10} \xi < -2.5$ is measured. A value of $4.27 \pm 0.04 \mbox{(stat.)} ^{+0.65}_{-0.58} 
\mbox{(syst.)}$ mb is extracted from the data. The total DD cross section integrated over the region $\Delta \eta > 3$, $M_X > 10$ GeV and $M_Y > 10$~GeV is also measured.
A value of $0.93 \pm 0.01 \mbox{(stat.)} ^{+0.26}_{-0.22} \mbox{(syst.)}$ mb is extracted from the data.

\section{Forward rapidity gap cross section}
\label{sec:fwdgap}

\begin{figure}[htb]
\centering
\includegraphics[scale=0.7]{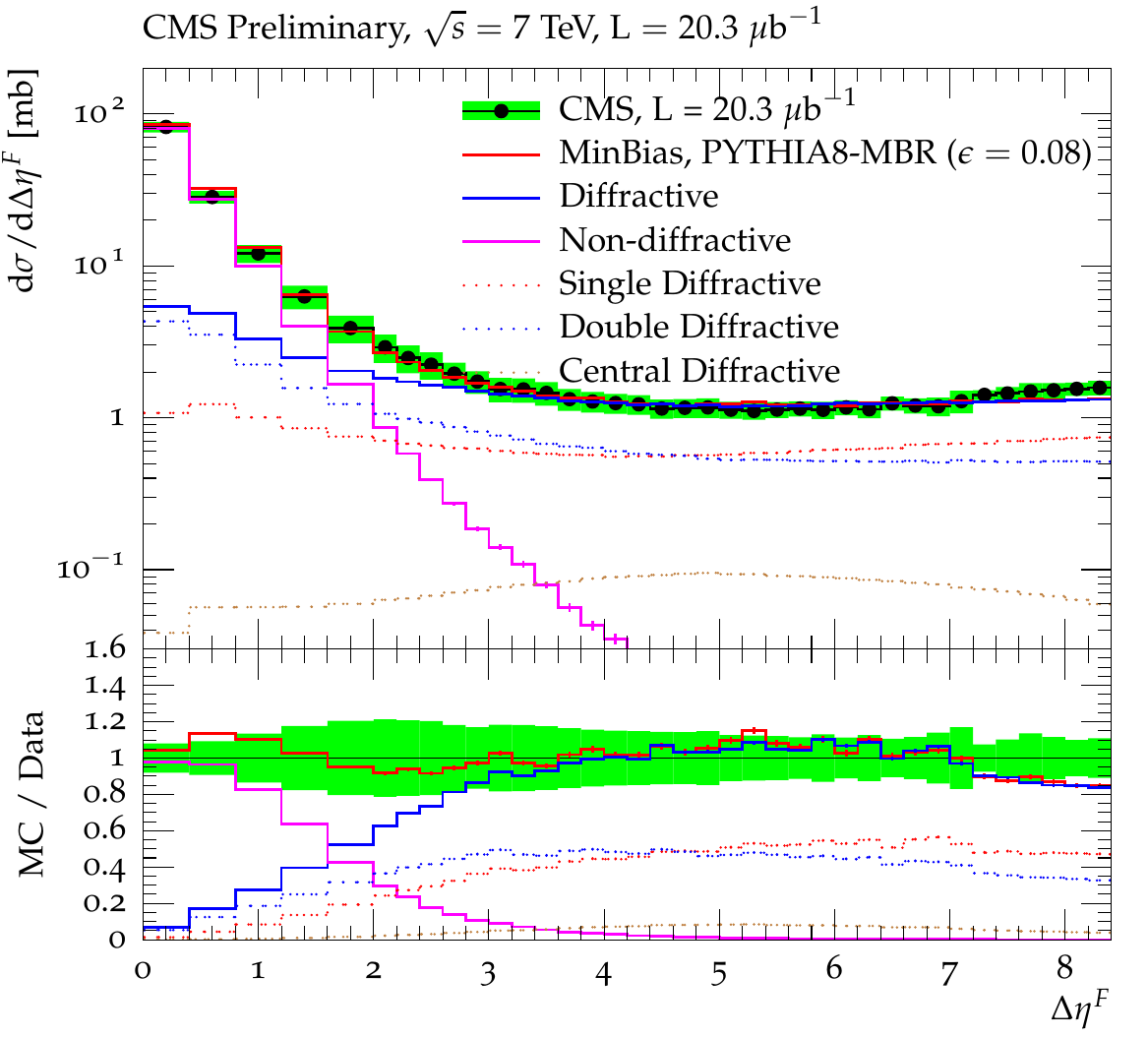}
\includegraphics[scale=0.7]{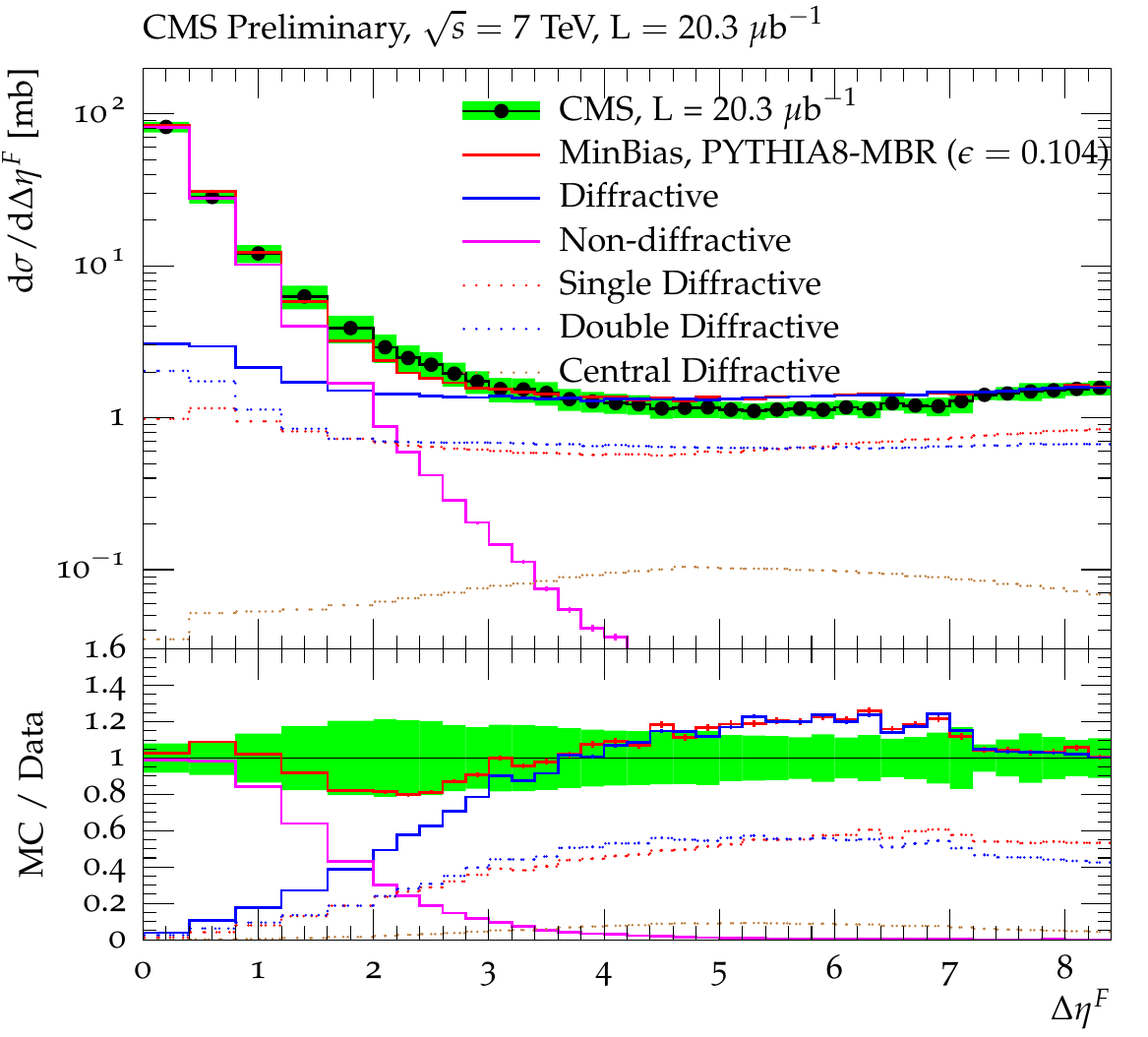}
\includegraphics[scale=0.7]{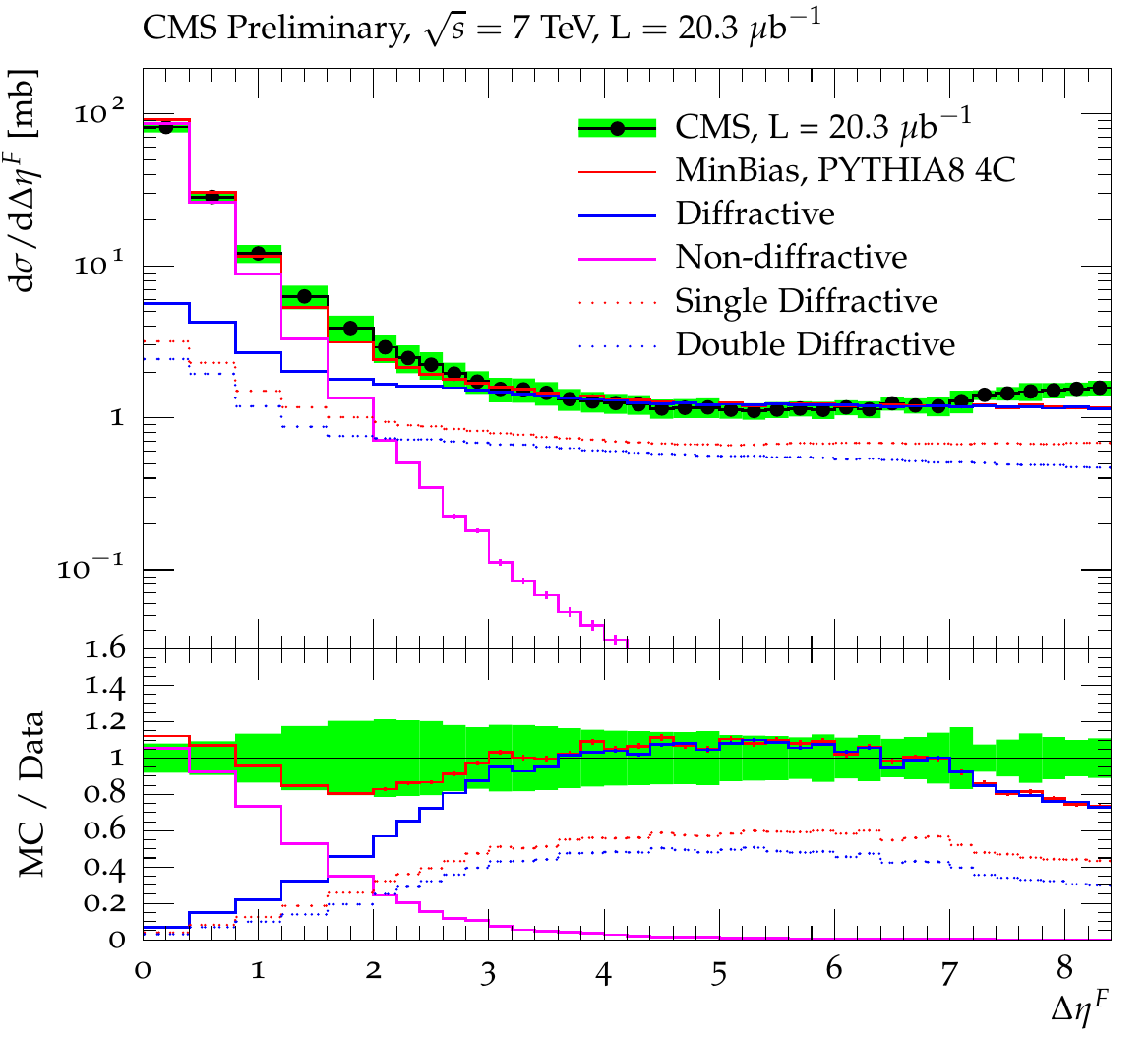}
\includegraphics[scale=0.7]{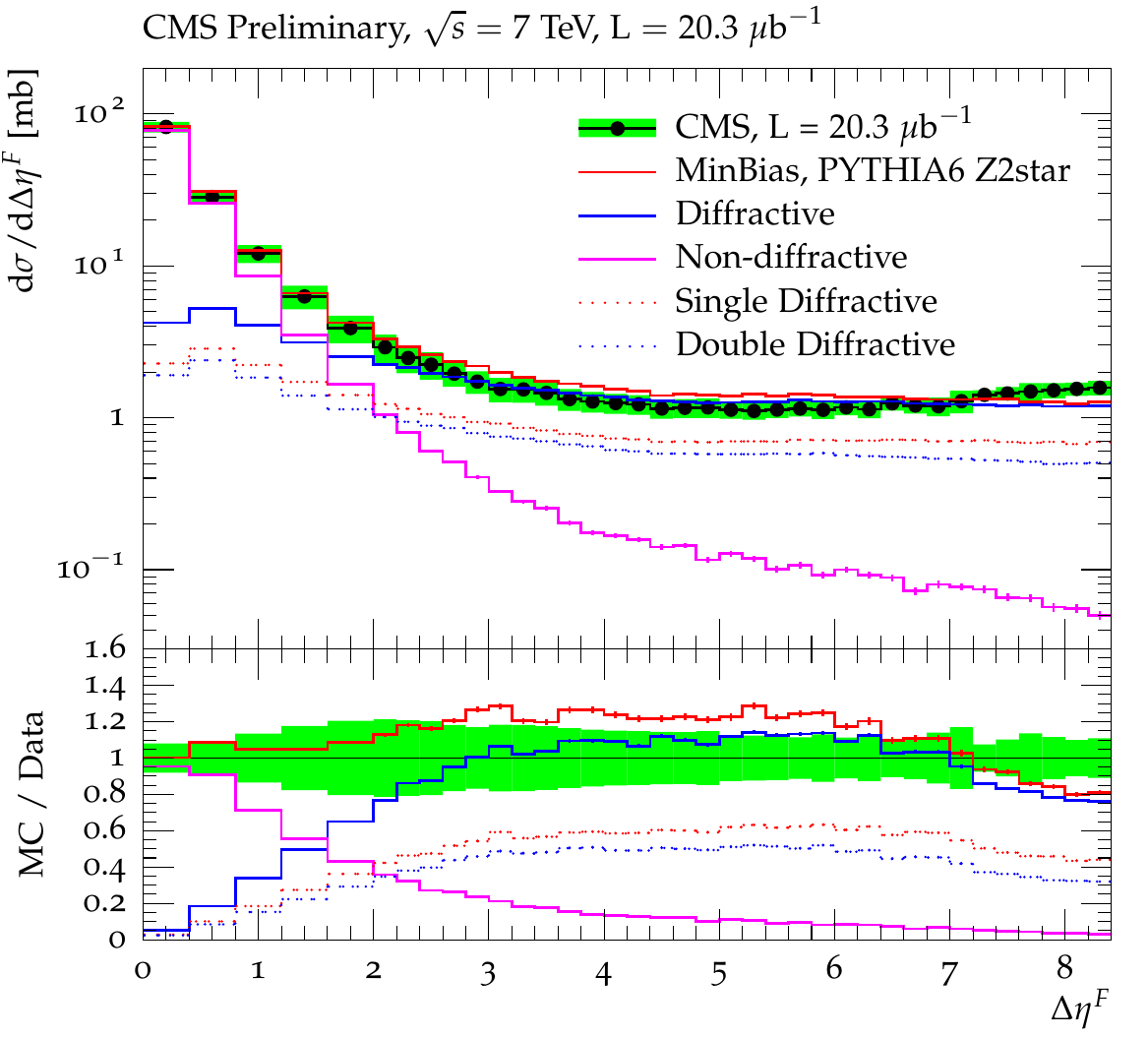}
\caption{The forward rapidity gap cross section $d\sigma/d\Delta \eta^F$ for events with at least one final state particle of transverse momentum $p_T > 200$ MeV in the region $|\eta|< 4.7$.
The measurement is compared to predictions of the MC event generators {\sc PYTHIA}8-MBR, {\sc PYTHIA}8-4C and {\sc PYTHIA}6-Z2$^*$. The bands represent the total systematic uncertainty.}
\label{fig:FRG}
\end{figure}

\begin{figure}[ht!]
\centering
\includegraphics[scale=0.7]{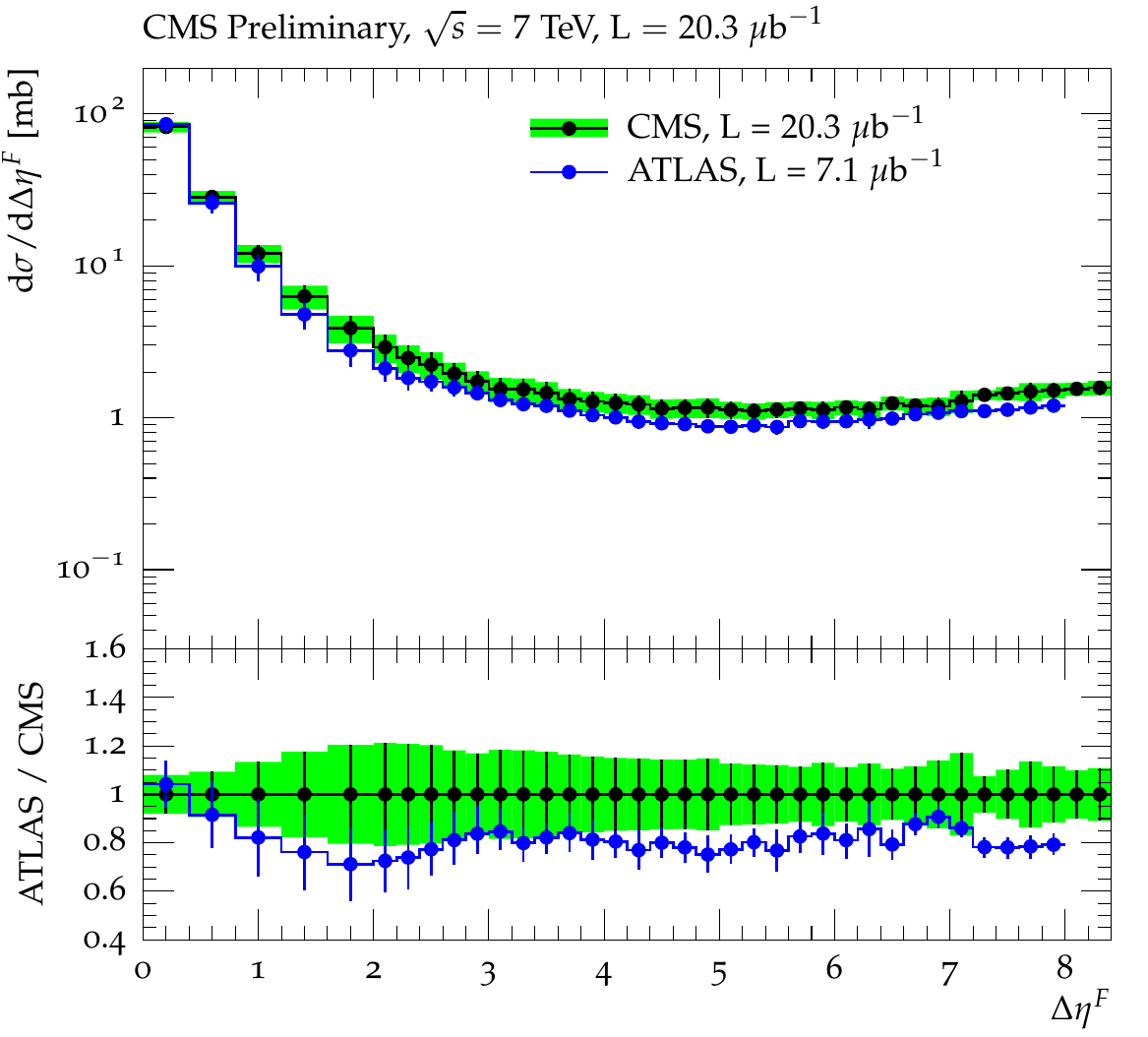}
\caption{Comparison of the CMS and ATLAS measurements of the forward rapidity gap cross section $d\sigma/d\Delta \eta^F$. The band represents the total systematic uncertainty of the CMS measurement,
while the total uncertainty of the ATLAS result is shown by the error bars.}  
\label{fig:CMSATLAS}
\end{figure}

The measurement of the forward rapidity gap cross section~\cite{CMS:2013mda} is based on a sample of $pp$ collisions at $\sqrt{s} = 7$ TeV selected online by a MB trigger.
The data sample corresponds to an integrated luminosity of 20.3 $\mu\mbox{b}^{-1}$ collected in a low pileup scenario with an average number of inelastic $pp$ collisions per bunch crossing of 0.007. 
The largest forward rapidity gap, $\Delta \eta^F$, is measured in the acceptance of the central CMS detector ($|\eta|< 4.7$) and defined as $\Delta \eta^F = max(4.7-\eta_{max},\eta_{min}+4.7)$.
The correction for the trigger efficiency and the subtraction of the beam induced background are determined in a data-driven way. The migration and detector acceptance corrections are evaluated
with an iterative Bayesian unfolding technique~\cite{Adye:2011gm}. The corrected results are defined for events with at least one final state particle of transverse momentum $p_T > 200$ MeV
in the region $|\eta|< 4.7$. The dominant sources of systematic uncertainties are the energy scale of the HF calorimeter and the modelling of hadronization and diffraction. The forward rapidity gap 
cross section is presented in figure~\ref{fig:FRG} and compared \\
 \\
to predictions of the MC event generators {\sc PYTHIA}8-MBR, {\sc PYTHIA}8-4C and {\sc PYTHIA}6 
with tune Z2$^*$~\cite{Chatrchyan:2013gfi}. The predictions from {\sc PYTHIA}8-MBR are shown for two values of the Pomeron intercept, $\alpha = 1.08$ and $\alpha = 1.104$. The bands represent the total 
systematic uncertainty of the order of $20\%$. The minimum-bias sample is dominated by ND events in the region of small forward rapidity gap $\Delta \eta^F < 3$. At larger $\Delta \eta^F$, the ND 
contribution is suppressed and the flattening of the exponential behaviour is due to the contribution of the diffractive events. Most of the predictions underestimate the cross section at large values 
of $\Delta \eta^F$. The prediction from {\sc PYTHIA}8-MBR with a Pomeron intercept $\alpha = 1.104$ gives the best description of the data at large values of $\Delta \eta^F$ but overestimates 
the cross section in the region of intermediate gap size.
A comparison of the CMS and ATLAS~\cite{Aad:2012pw} measurements is shown in figure~\ref{fig:CMSATLAS}. Except for the pseudorapidity coverage, which is slightly different between the CMS ($|\eta|< 4.7$)
and ATLAS ($|\eta|< 4.9$) results, the hadron-level definitions of the cross sections are identical. The CMS result is consistent with the result of the ATLAS collaboration and extends the ATLAS
measurement by 0.4 unit of gap size.
 
\section{Diffractive dijet cross section}
\label{sec:dijet}

The observation of a diffractive component to the dijet production cross section~\cite{Chatrchyan:2012vc} is based on a sample of $pp$ collisions at $\sqrt{s} = 7$ TeV selected online by a single jet 
trigger with a $p_T$ threshold of 6 GeV. The data sample was collected when the LHC was operating at low instantaneous luminosity and corresponds to an integrated luminosity of 2.7 n$\mbox{b}^{-1}$, 
with an average pileup of 0.09. This sample is dominated by nondiffractive inclusive dijet events and the selection of a diffractive-enhanced sample is based on the presence of a LRG in the event. 
The cross section is measured as a function of the variable $\tilde{\xi}$ that coincides with the fractional momentum loss of the scattered proton, $\xi$, for SD events. \\

The following requirements are applied offline. Events with at least two jets with transverse momentum $p_T > 20$ GeV and $|\eta| < 4.4$ are selected. Jets are reconstructed from particle candidates 
with the anti-$k_T$ jet clustering algorithm~\cite{Cacciari:2008gp} with a distance parameter $R = 0.5$. At least one primary vertex is required to be present within 24 cm of the nominal interaction point 
along the beam direction. The diffractive-enhanced sample is defined by the selections $\eta_{max} < 3$ and $\eta_{min} > -3$, respectively. This condition corresponds to the presence of a pseudorapidity 
gap of at least 1.9 units within the acceptance of the detector, $|\eta| < 4.9$. The variable $\tilde{\xi}$ is reconstructed from the energy $E^i$ and the longitudinal momentum $p_z^i$ of the particle 
candidates measured in the central detector, according to the relation $\tilde{\xi}^\pm = \sum(E^i \pm p_z^i)/\sqrt{s}$. The variables $\tilde{\xi}^+$ and $\tilde{\xi}^-$ are associated to events in which 
the dissociated system is measured on the negative and positive $z$-side of CMS, respectively. \\ 

\begin{figure}[ht!]
\centering
\includegraphics[scale=0.4]{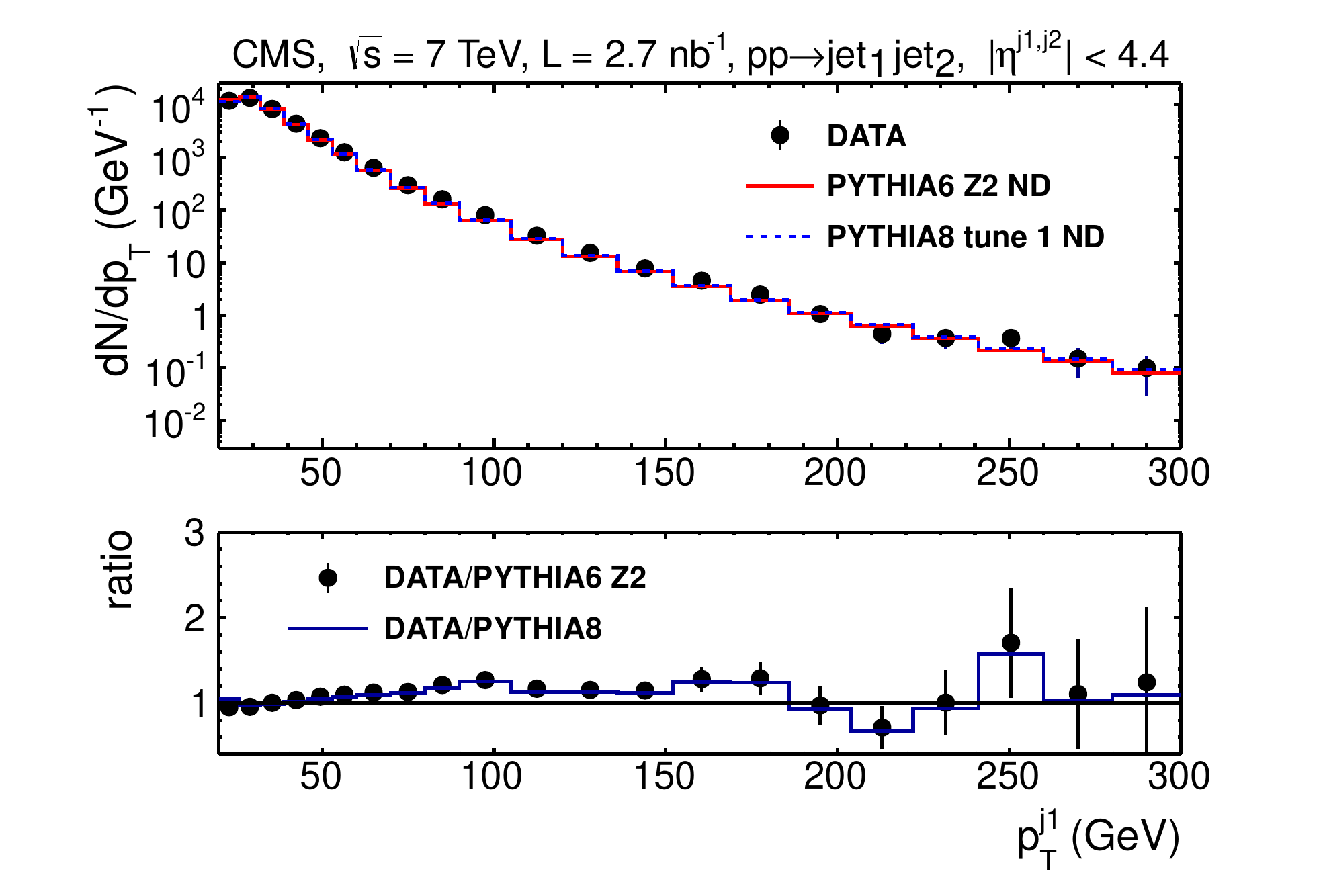}
\includegraphics[scale=0.4]{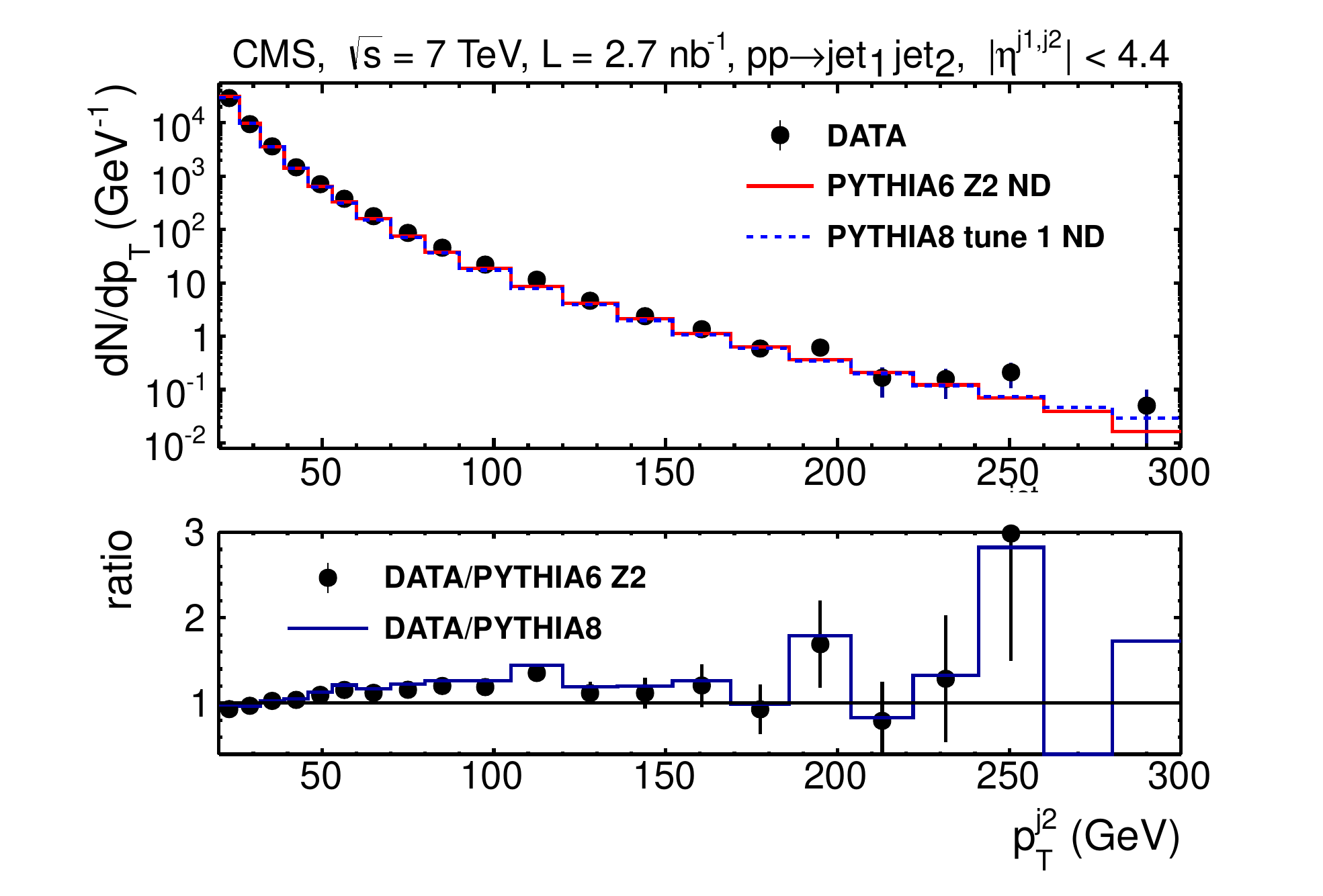}
\includegraphics[scale=0.4]{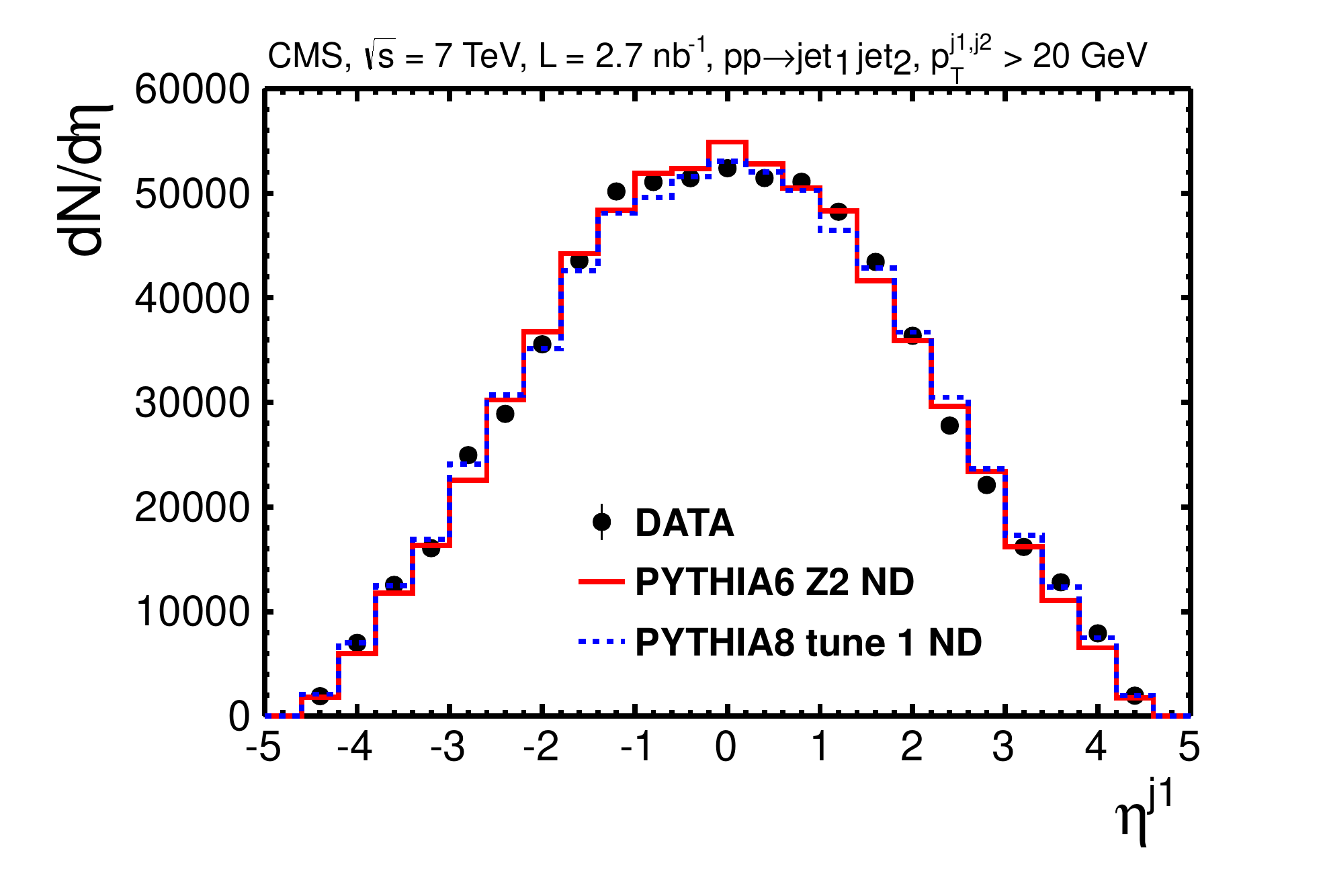}
\includegraphics[scale=0.4]{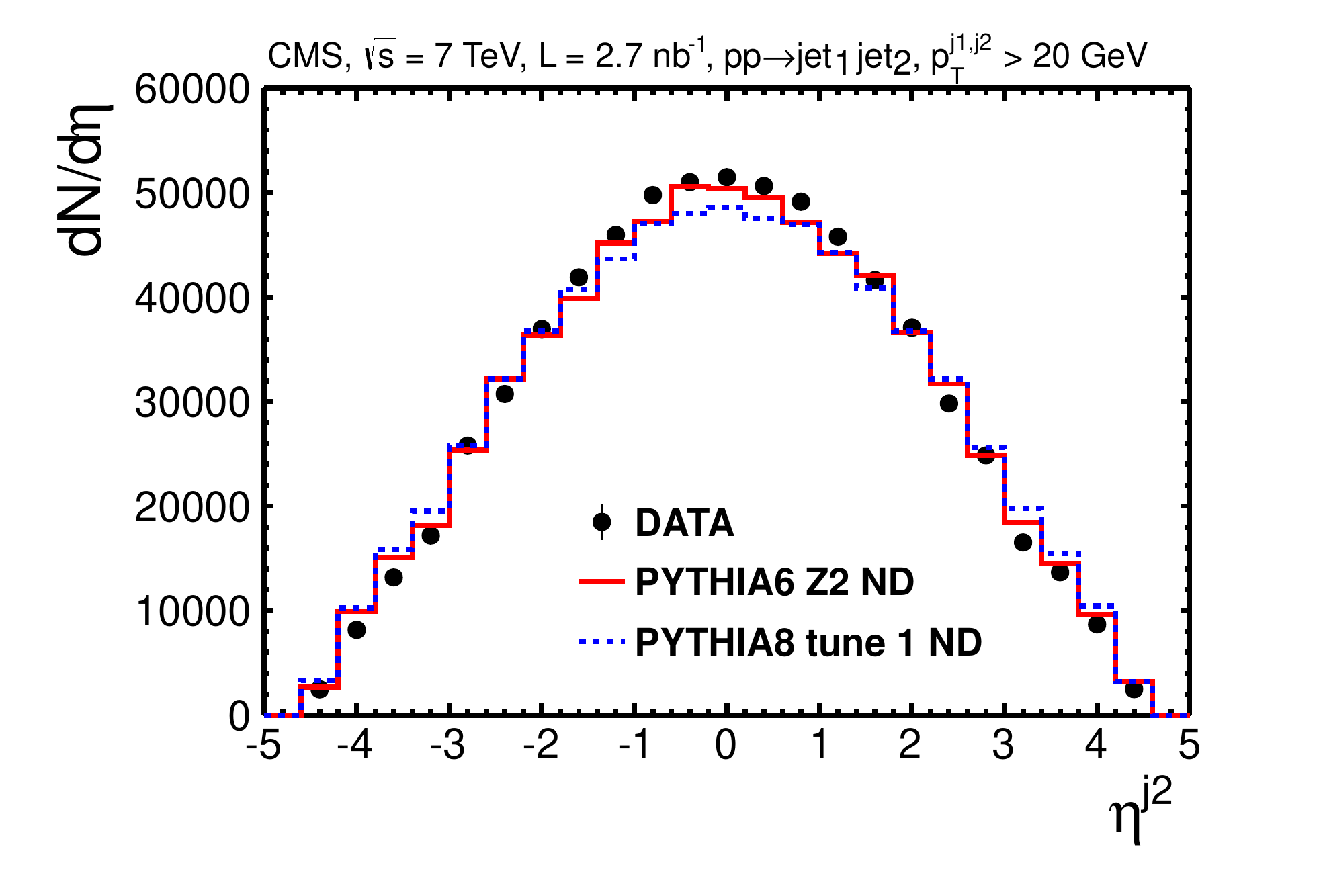}
\caption{Detector-level distributions of the transverse momentum and pseudorapidity of the leading and second-leading jets. The data are compared to predictions of the MC event generators
{\sc PYTHIA}6-Z2 and {\sc PYTHIA}8 tune 1. The MC predictions are given for the nondiffractive component only. The error bars indicate the statistical uncertainty. The MC predictions are normalized 
to the number of events in the data.}
\label{fig:dijet}
\end{figure}

The distributions of the transverse momentum and pseudorapidity of the leading and second-leading jets are presented in figure~\ref{fig:dijet} and compared to predictions of the MC event generators
{\sc PYTHIA}6 with tune Z2~\cite{Field:2010bc} and {\sc PYTHIA}8 with tune 1~\cite{Sjostrand:2007gs}. The MC predictions are given for the nondiffractive component only. Both {\sc PYTHIA}6-Z2 and 
{\sc PYTHIA}8 tune 1 describe the transverse momentum distributions equally well, while the pseudorapidity distributions are better described by the predictions of {\sc PYTHIA}6-Z2. 

\begin{figure}[ht!]
\centering
\vspace*{-0.4cm}  
\includegraphics[scale=0.7]{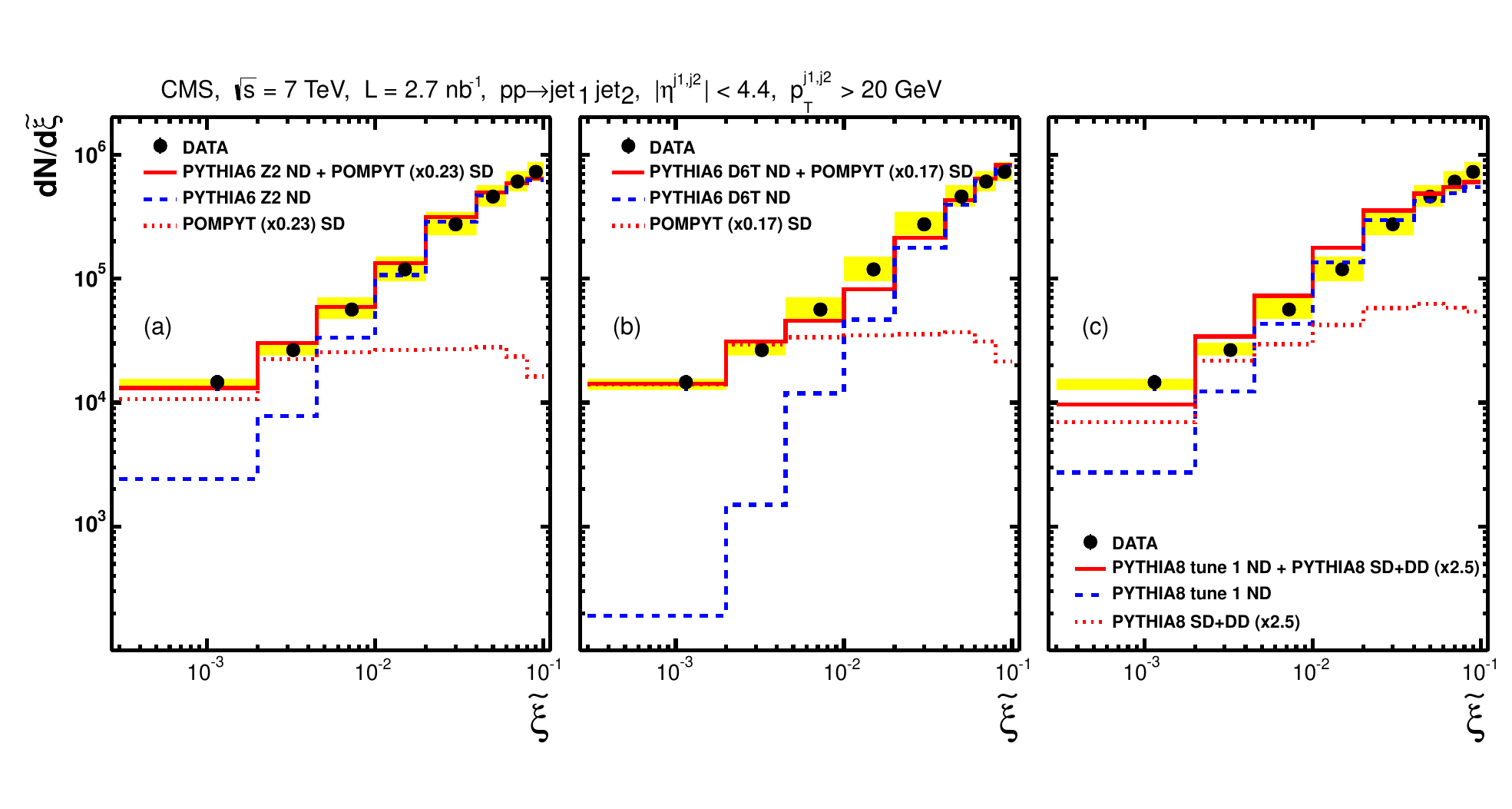}
\vspace*{-0.7cm}
\caption{Detector-level $\tilde{\xi}$ distribution. The data are compared to the combinations of different MC event generators. The predictions of {\sc PYTHIA}6-Z2 + {\sc POMPYT} (a),  
{\sc PYTHIA}6-D6T + {\sc POMPYT} (b) and {\sc PYTHIA}8 tune 1 (c) are shown. The relative diffractive contributions are scaled by the values given in the legend. The error bars indicate the statistical
uncertainty, and the band represents the uncertainty on the calorimeter energy scale. The sum of the MC predictions is normalized to the number of events in the data.}
\label{fig:xibefore}
\end{figure}

The distribution of the variable $\tilde{\xi}$ is presented in figure~\ref{fig:xibefore} before the $\eta_{max}$ or $\eta_{min}$ selection and compared to predictions of different MC event generators.
The data are described by a combination of diffractive and nondiffractive contributions. The relative diffractive component and the overall normalisation factor are determined by minimising the
difference between the distribution of the data and the sum of the nondiffractive and diffractive MC predictions. The data are compared in figure~\ref{fig:xibefore}(a) to a combination of the MC event 
generators {\sc PYTHIA}6-Z2 and {\sc POMPYT}~\cite{Bruni:1993is} that describe the nondiffractive and single diffractive components, respectively. According to the minimization procedure, 
the predictions of {\sc POMPYT} need to be scaled by a factor 0.23 to describe the data. Figure~\ref{fig:xibefore}(b) compares the data to a combination of {\sc PYTHIA}6-D6T and {\sc POMPYT}. 
The scale factor to be applied to the {\sc POMPYT} normalisation has a value of 0.17 for this combination. The data are compared in figure~\ref{fig:xibefore}(c) to the predictions of {\sc PYTHIA}8 
tune 1. The ND, SD and DD contributions are all simulated by {\sc PYTHIA}8, and the diffractive components need to be scaled by a factor 2.5 in that case. The diffractive generators {\sc POMPYT}  
and {\sc PYTHIA}8 use the dPDFs determined by the H1 collaboration (H1 fit B) ~\cite{Aktas:2006hy}. The parametrisation of the 
Pomeron flux in {\sc POMPYT} is also based on the QCD fits to the HERA data~\cite{Aktas:2006hy}, while {\sc PYTHIA}8 uses a different normalisation of the Pomeron flux~\cite{Navin:2010kk}. The large 
difference between the scale factors in {\sc POMPYT} and {\sc PYTHIA}8 is a consequence of this difference in the implementation of the Pomeron flux. None of the diffractive generators include 
the presence of the rapidity gap survival probability. In the case of {\sc POMPYT}, the scale factor to be applied is a consequence of this fact. In the case of {\sc PYTHIA}8, 
the normalisation of the Pomeron flux is about a factor 10 higher than in {\sc POMPYT} in the low-mass region. Taking this difference into account, the scale factor to be applied to the diffractive 
components in {\sc PYTHIA}8 would have a value of about 0.25. The uncertainty of the scale factors is estimated by changing the fitting procedure and is found to be $\sim 20 \%$. The three different 
values are therefore compatible within the uncertainty. 

\begin{figure}[ht!]
\centering
\includegraphics[scale=0.4]{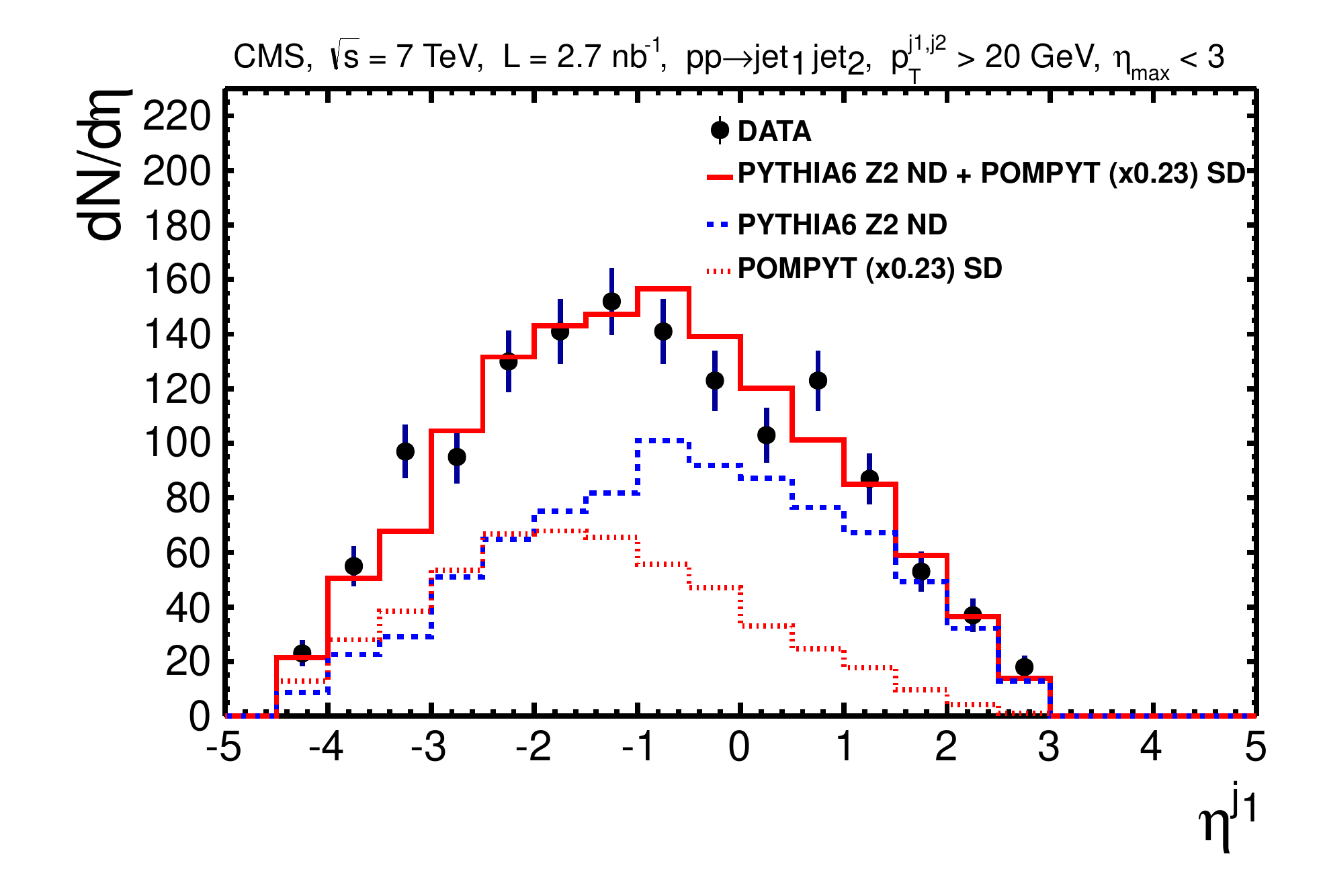}
\includegraphics[scale=0.4]{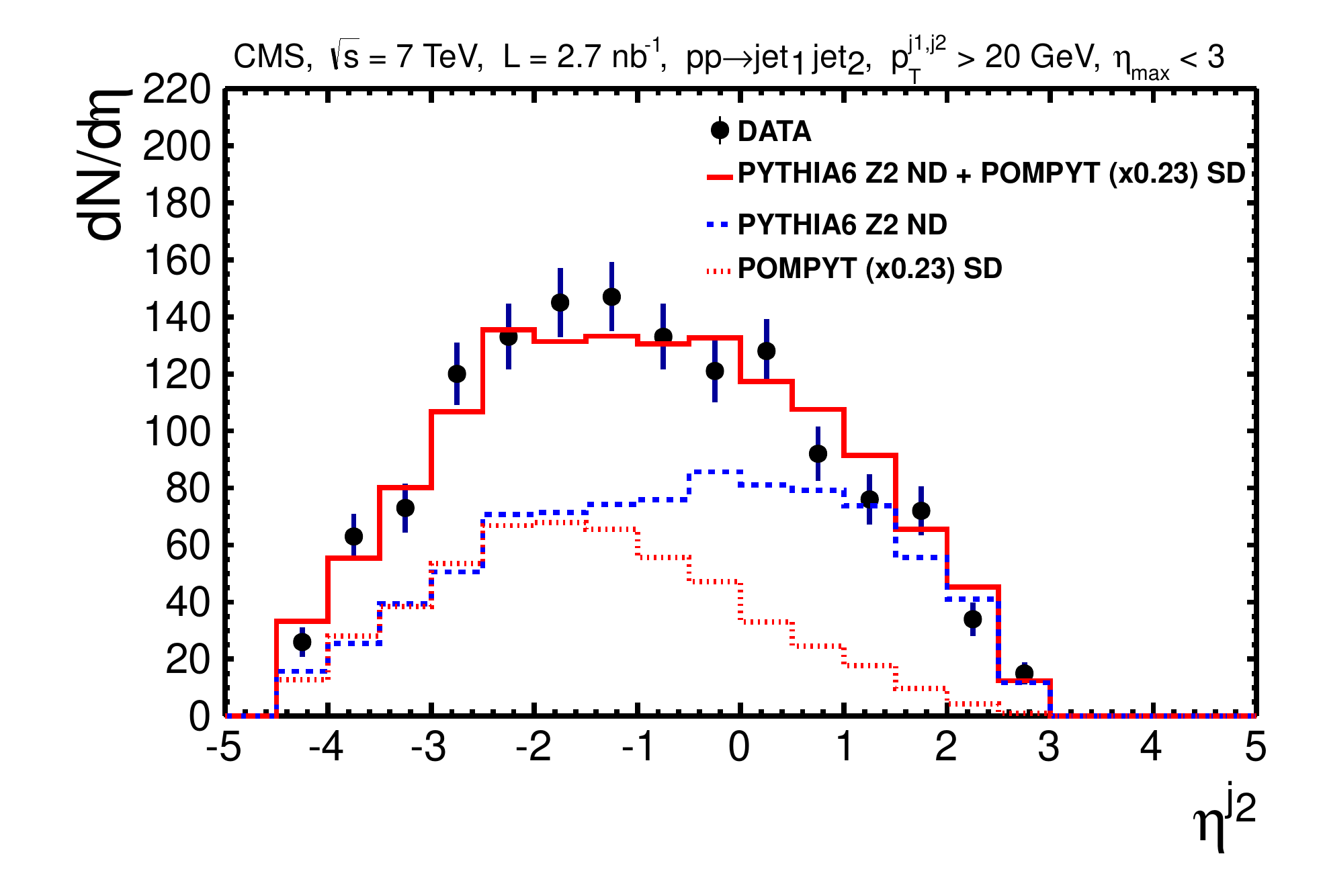}
\vspace*{-0.2cm}
\caption{Detector-level pseudorapidity distributions of the leading and second-leading jets after the selection $\eta_{max} < 3$ or $\eta_{min} > -3$. The data are compared to a combination of 
{\sc PYTHIA}6-Z2 and {\sc POMPYT} that describe the nondiffractive and single diffractive components, respectively. The individual MC predictions are also shown. The relative diffractive 
contribution is scaled by a factor 0.23. The error bars indicate the statistical uncertainty. The sum of the MC predictions is normalized to the number of events in the data.}
\label{fig:dijetafter}
\end{figure}

The pseudorapidity distributions of the two leading jets after the selection $\eta_{max} < 3$ or $\eta_{min} > -3$ are presented in figure~\ref{fig:dijetafter}. The requirement of the pseudorapidity gap
enhances the diffractive component in the data. The asymmetry observed in the pseudorapidity distributions is a consequence of the diffractive topology, the jets being mainly produced in the 
hemisphere opposite to that of the gap in SD events. The data are compared to a combination of the MC event generators {\sc PYTHIA}6-Z2 and {\sc POMPYT}, with the normalisation of the diffractive
component scaled by a factor 0.23 as determined by the fitting procedure previously described.

\begin{figure}[ht!]
\centering
\includegraphics[scale=0.5]{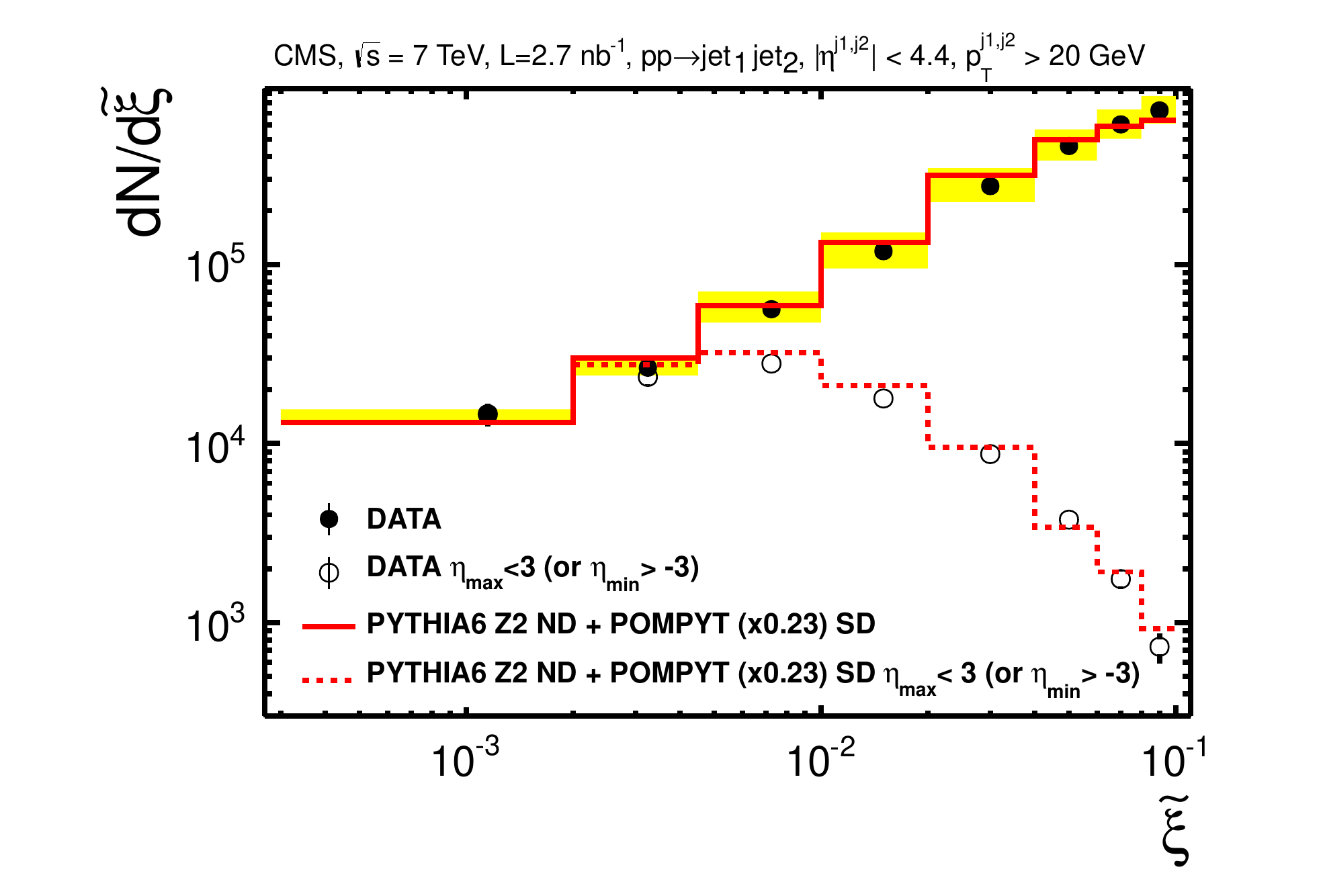}
\caption{Detector-level $\tilde{\xi}$ distribution before and after the rapidity gap selection $\eta_{max} < 3$ or $\eta_{min} > -3$. The data are compared to a combination of {\sc PYTHIA}6-Z2 
and {\sc POMPYT}. The relative diffractive contribution is scaled by a factor 0.23. The error bars indicate the statistical uncertainty, and the band represents the calorimeter energy scale uncertainty. 
The sum of the MC predictions is normalized to the number of events in the data.}
\label{fig:xiafter}
\end{figure}

The distribution of the variable $\tilde{\xi}$ is presented in figure~\ref{fig:xiafter} before and after the selection $\eta_{max} < 3$ or $\eta_{min} > -3$. The data are compared to a combination of 
{\sc PYTHIA}6-Z2 and {\sc POMPYT}, with the normalisation of {\sc POMPYT} scaled by 0.23. The requirement of the pseudorapidity gap mainly rejects events in the region of high $\tilde{\xi}$ values 
dominated by the ND contribution. The region of low $\tilde{\xi}$ values where the diffractive contributions dominate is only slightly affected by the selection. \\

\begin{figure}[ht!]
\centering
\vspace*{0.8cm}
\includegraphics[scale=0.5]{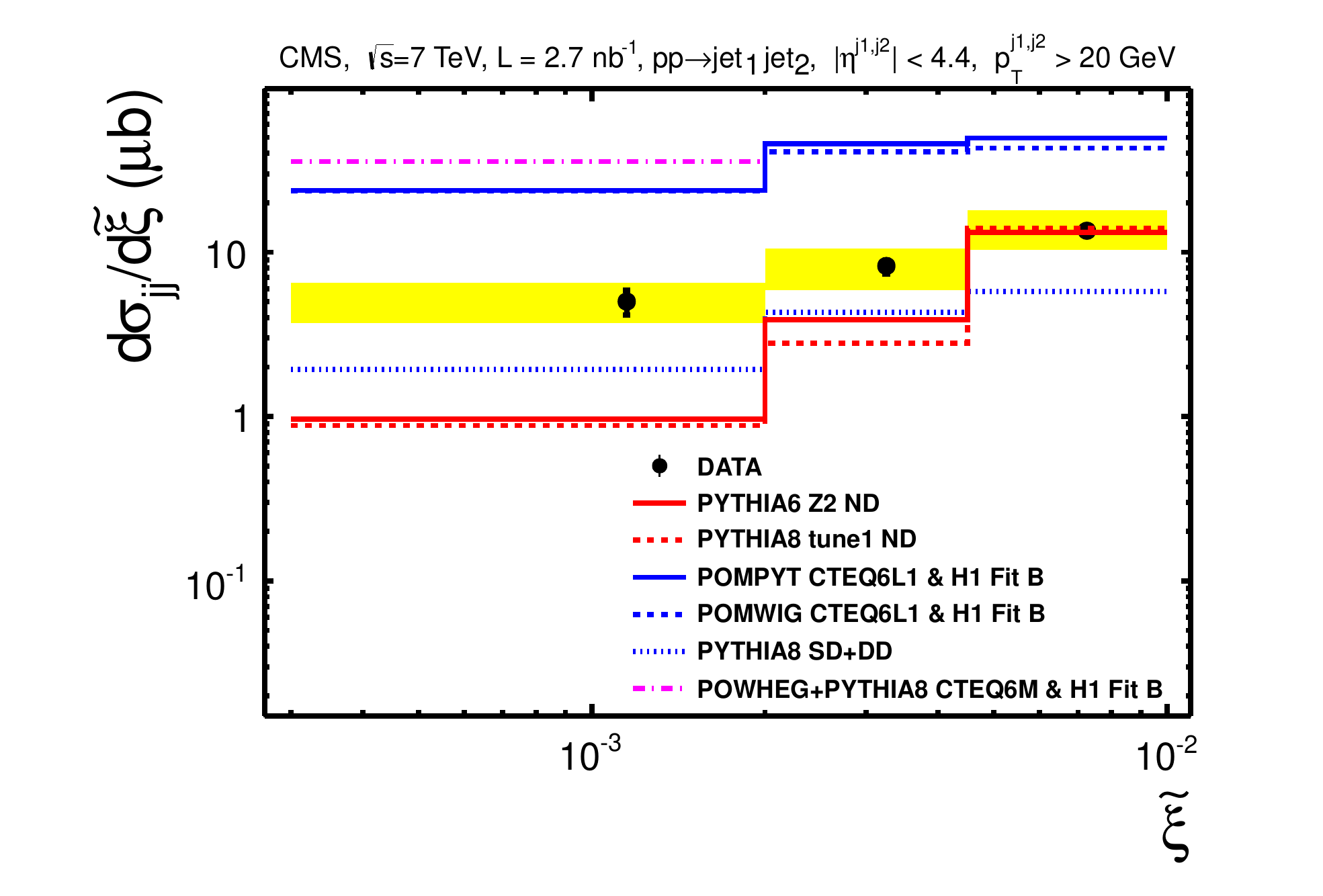}
\caption{Differential cross section for inclusive dijet production as a function of $\tilde{\xi}$ for jets with transverse momentum $p_T > 20$ GeV and $|\eta| < 4.4$. The error bars indicate 
the statistical uncertainty and the band represents the systematic uncertainties added in quadrature. The data are compared to predictions of LO MC event generators,
{\sc PYTHIA}6-Z2 and {\sc PYTHIA}8 tune 1 for the ND contribution, and {\sc POMPYT}, {\sc POMWIG} and {\sc PYTHIA}8 tune 1 for the diffractive contribution. The generators {\sc POMPYT} 
and {\sc POMWIG} simulate only the SD component, while {\sc PYTHIA}8 simulates both the SD and DD components. The NLO prediction based on {\sc POWHEG} is also shown in the lowest $\tilde{\xi}$ bin only.}
\label{fig:finalxs}
\end{figure}

The differential cross section for inclusive dijet production as a function of $\tilde{\xi}$ is corrected for the detector acceptance and the migration of reconstructed variables by a bin-by-bin 
unfolding procedure based on a combination of {\sc PYTHIA}6-Z2 and {\sc POMPYT}, with the normalisation of {\sc POMPYT} scaled by 0.23. The bin-by-bin corrections are found 
to be consistent with that obtained with the SVD~\cite{Hocker:1995kb} and Bayesian~\cite{D'Agostini:1994zf} unfolding techniques. The data are also corrected for the trigger efficiency and the effect 
of pileup. The total systematic uncertainty of the measurement is $\sim 30\%$ and dominated by the uncertainty on the jet energy scale. The differential cross section is presented 
in figure~\ref{fig:finalxs} for jets with transverse momentum $p_T > 20$ GeV and $|\eta| < 4.4$. The data are compared to predictions of leading-order (LO) MC event generators, {\sc PYTHIA}6-Z2 and 
{\sc PYTHIA}8 tune 1 for the ND contribution, and {\sc POMPYT}, {\sc POMWIG}~\cite{Cox:2000jt} and {\sc PYTHIA}8 tune 1 for the diffractive contribution. The generators {\sc POMPYT} and {\sc POMWIG} 
simulate only the SD component, while {\sc PYTHIA}8 simulates both the SD and DD components. The next-to-leading-order (NLO) prediction based on {\sc POWHEG}~\cite{Alioli:2010xa} is also shown 
in the lowest $\tilde{\xi}$ bin only. \\

The generators {\sc PYTHIA}8, {\sc POMPYT} and {\sc POMWIG}, and the NLO predictions based on the {\sc POWHEG} framework use the dPDFs determined by the H1 collaboration 
(H1 fit B~\cite{Aktas:2006hy}). The main difference between {\sc POMPYT} and {\sc POMWIG} is that {\sc POMPYT} uses the {\sc PYTHIA} framework for the hadronization, while {\sc POMWIG} is based on 
the {\sc HERWIG}~\cite{Corcella:2000bw} framework. The predictions of {\sc PYTHIA}6-Z2 and {\sc PYTHIA}8 tune 1 for the ND contribution underestimate the data in the low $\tilde{\xi}$ region. 
The predictions of {\sc POMPYT}, {\sc POMWIG} and {\sc POWHEG} are needed to describe the low $\tilde{\xi}$ region but overestimate the data by a factor $\sim 5$. The ratio of the measured cross section 
to that expected from the diffractive MC models gives an upper limit of the rapidity gap survival probability, the measured cross section including a proton dissociation contribution that needs 
to be subtracted to get an estimate of the survival probability. For {\sc POMPYT} and {\sc POMWIG}, the value of the upper limit is $0.21 \pm 0.07$. After correction for the proton dissociation 
contribution present in the measured cross section and in the dPDFs, an estimate of the rapidity gap survival probability can be obtained and is found to be $0.12 \pm 0.05$ for the LO MC generators 
{\sc POMPYT} and {\sc POMWIG}. If the NLO predictions based on {\sc POWHEG} are used, the estimation of the rapidity gap survival probability gives a lower value of $0.08 \pm 0.04$. 
It should be mentioned that the dPDFs and the proton dissociation contribution are poorly constrained in the region of the measurement. This contribution to the uncertainty of the survival probability
is not taken into account. The normalisation of the diffractive predictions of {\sc PYTHIA}8 disagrees with that of {\sc POMPYT} and {\sc POMWIG}, and the predictions would need to be 
scaled by a factor $\sim 2.5$ to describe the data. This is a consequence of the different normalisation of the Pomeron flux in {\sc PYTHIA}8 with respect to that used in the H1 fit B. 
For that reason, {\sc PYTHIA}8 can not directly be used to estimate the gap survival probability. The present value is compatible with the CDF measurement~\cite{Affolder:2000vb,Aaltonen:2012tha}.

\section{$W$ and $Z$ boson events with a LRG signature}
\label{sec:WZ}

The observation of $W$ and $Z$ boson events with a large pseudorapidity gap~\cite{Chatrchyan:2011wb} is based on a sample of $pp$ collisions at $\sqrt{s} = 7$ TeV selected online by requiring a high 
transverse momentum electron or muon with $p_T$ thresholds varying between 10 and 17 GeV for electrons and between 9 and 15 GeV for muons. The data sample was collected when the LHC was steadily 
increasing the instantaneous luminosity and corresponds to an integrated luminosity of 36 p$\mbox{b}^{-1}$. The average pileup is increasing from less than one additional inelastic $pp$ collision per bunch 
crossing to more than two. This sample is dominated by inclusive $W$ and $Z$ boson events and the selection of a diffractive-enhanced sample is based on the presence of a LRG in the event. \\

The identification of $W$ and $Z$ bosons is based on the presence of isolated electrons and muons with high transverse momentum. The following requirements are applied offline. In order to limit the 
effects of pileup, events with more than one vertex are rejected. Isolated electron and muon candidates are selected with a transverse momentum $p_T > 25$ GeV and $|\eta| < 2.5$. Electrons from photon 
conversions are rejected. Muons are required to be identified by two different reconstruction algorithms~\cite{Khachatryan:2010xn}. Jets and missing transverse momenta are determined from particle 
candidates. For the jet reconstruction, the anti-$k_T$ jet clustering algorithm~\cite{Cacciari:2008gp} is used with a distance parameter $R = 0.5$. Jets are selected with a transverse momentum 
$p_T > 30$ GeV and $|\eta| < 2.5$. 

An event is selected as a $W \to l \nu$ candidate if it fulfills the following requirements. One isolated electron or muon with $p_T > 25$ GeV and $|\eta| < 1.4$ is present in the event, and no other
isolated electron or muon with $p_T > 10$ GeV is found. The missing transverse momentum is greater than 30 GeV and the transverse mass of the charged lepton and the neutrino greater than 60 GeV. 
An event is selected as a $Z \to l l$ candidate if the following requirements are satisfied. Two isolated electrons or muons with opposite charge and $p_T > 25$ GeV are present in the event. At least
one of the leptons has $|\eta| < 1.4$ and the invariant mass of the dilepton system is between 60 and 120 GeV. \\

The presence of a LRG is defined by the absence of activity above the noise threshold in at least one of the HF calorimeters. This condition defines the diffractive-enhanced samples of $W$ and $Z$ 
boson events and corresponds to the presence of a pseudorapidity gap of at least 1.9 units in the event, the fiducial coverage of the HF calorimeters being limited to the region $3 < |\eta| < 4.9$. 
Soft pileup events with few or no activity in the central region of the detector are not rejected by the single vertex requirement and have the ability to fill the pseudorapidity gap in the forward region.
The contribution of these events to the energy measured in the HF calorimeters is found to be well reproduced by the CMS simulation. \\

Diffractively produced $W$ and $Z$ bosons are more likely to be found in the hemisphere opposite to that of the gap. This is a consequence of the fact that dPDFs peak at smaller $x$ values   
than the inclusive parton distribution functions (PDFs). The boson produced being boosted in the direction of the parton with the highest momentum fraction, it is more likely to be found in the 
direction of the dissociated system, opposite to that of the gap. This property is used to estimate the diffractive component in the samples of $W$ and $Z$ boson events with a LRG. \\

\begin{figure}[ht!]
\centering
\vspace*{-0.2cm}
\includegraphics[scale=0.45]{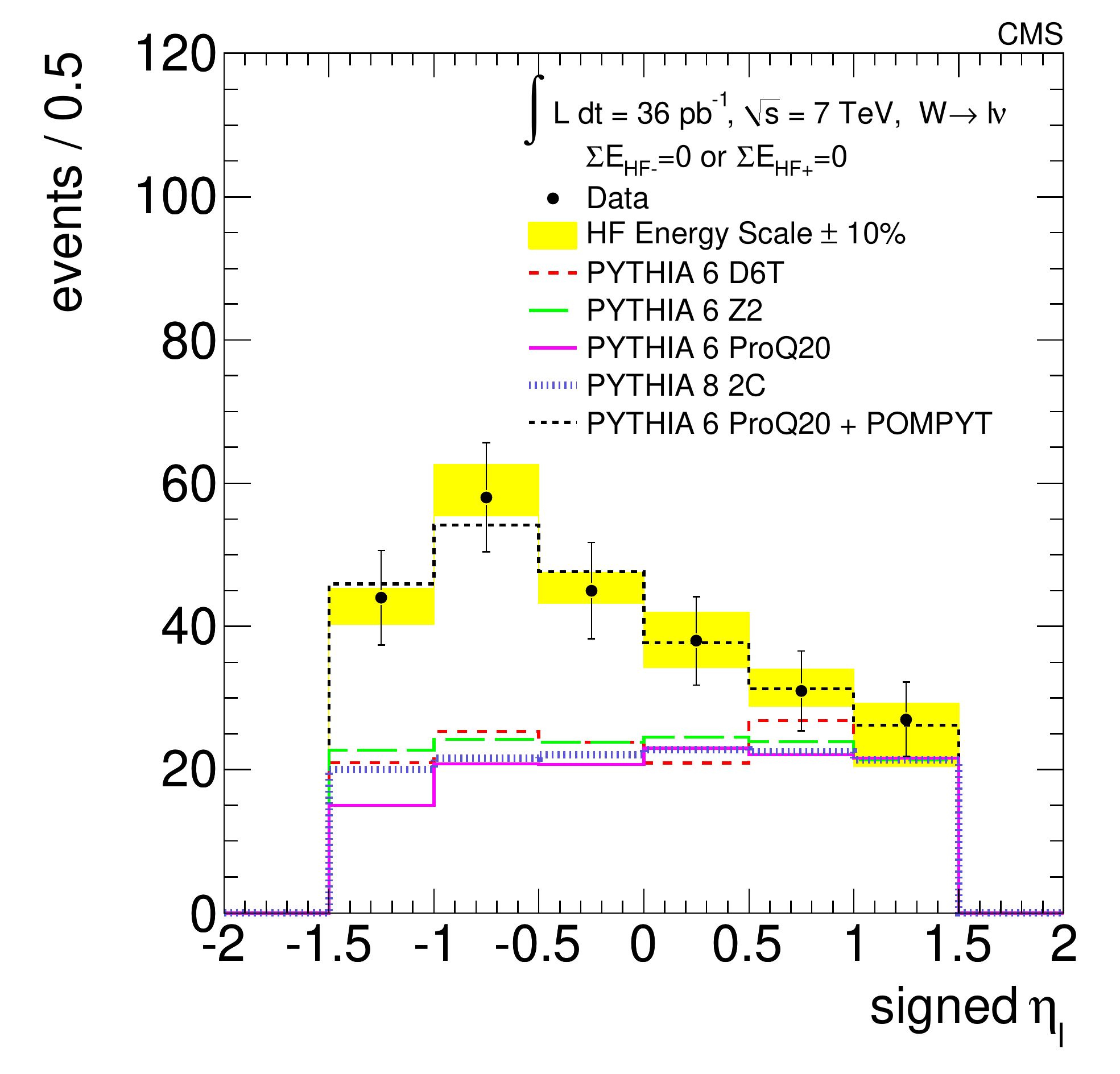}
\vspace*{-0.3cm}
\caption{Distribution of the signed pseudorapidity of the charged lepton, $\eta_l$, in $W$ events with a LRG for the combination of the electron and muon channels. The pseudorapidity is positive 
when the gap and the lepton are in the same hemisphere and negative otherwise. The data are compared to a combination of {\sc PYTHIA}6 Pro-Q20 and {\sc POMPYT} that describe the nondiffractive and 
single diffractive components, respectively. The nondiffractive predictions are also shown for the MC generators {\sc PYTHIA}8 2C and {\sc PYTHIA}6 with tunes D6T, Z2 and Pro-Q20. The band represents 
the energy scale uncertainty of the HF calorimeters.}
\label{fig:WLRG}
\end{figure}

The distribution of the signed pseudorapidity of the charged lepton, $\eta_l$, in $W$ events with a LRG is shown in figure~\ref{fig:WLRG} for the combination of the electron and muon channels. The sign 
is defined according to the hemispheres in which the pseudorapidity gap and the charged lepton are found. The pseudorapidity is positive when the gap and the lepton are in the same hemisphere and
negative otherwise. A total of 147 events are found with the charged lepton in the hemisphere opposite to that of the gap, and 96 events with the lepton in the same hemisphere. The asymmetry can be defined
as the ratio of the difference between the numbers of LRG events with positive and negative $\eta_l$ and the total number of events. The corresponding asymmetry is $21.0 \pm 6.4 \%$. For the sample of
$Z$ bosons with a LRG, a compatible asymmetry of $20 \pm 16 \%$ is found. The data are compared to a combination of the MC event generators {\sc PYTHIA}6 Pro-Q20~\cite{Buckley:2009bj} and {\sc POMPYT} 
that describe the nondiffractive and single diffractive components, respectively. The predictions of {\sc POMPYT} show a strong asymmetry in  $\eta_l$, while the nondiffractive contribution estimated with 
{\sc PYTHIA}6 Pro-Q20 is symmetric. The relative diffractive contribution is determined by minimising the difference between the distribution of the data and the sum of the nondiffractive and diffractive 
MC predictions. According to the minimization procedure, the fraction of diffractive events in the sample of $W$ events with a LRG is found to be 
$(50.0 \pm 9.3 \, (\mbox{stat.}) \pm 5.2 \, (\mbox{syst.})) \%$. The choice of {\sc PYTHIA}6 Pro-Q20 to describe the nondiffractive component comes from the fact that this generator gives the best 
description of the energy distribution in the HF calorimeters, especially at low energy. The minimization procedures based on the combination of {\sc POMPYT} with {\sc PYTHIA}6-D6T, {\sc PYTHIA}6-Z2 
or {\sc PYTHIA}8 2C~\cite{Corke:2010yf} give similar results. The nondiffractive predictions of these generators are shown in figure~\ref{fig:WLRG}. The systematic uncertainty of $5.2 \%$ 
is estimated from the uncertainty on the HF calorimeter energy scale and the choice of different nondiffractive predictions in the fits.

\section{Pseudorapidity distribution of charged particles in a single diffractive enhanced sample}
\label{sec:eta}

The measurement of pseudorapidity distributions of charged particles~\cite{Chatrchyan:2014qka} is based on a sample of $pp$ collisions at $\sqrt{s} = 8$ TeV collected in July 2012 by a minimum-bias trigger
during a dedicated run, when the LHC was operating in a very low pileup scenario and with a non-standard $\beta^* = 90$ m optics configuration, with $\beta^*$ the amplitude function of the beam at the IP. 
These data correspond to an integrated luminosity of 45 $\mu\mbox{b}^{-1}$ with an average number of inelastic $pp$ collisions per bunch crossing of 0.04. The minimum-bias trigger is provided by the TOTEM
T2 telescopes. It contributes to the CMS global trigger decision and initiates the simultaneous readout of the CMS and TOTEM detectors. The trigger requires the presence of at least one track candidate
in one of the T2 telescopes in either side of the IP. Events reconstructed by the CMS and TOTEM detectors are combined offline. \\

The pseudorapidity distributions of charged particles $\textrm{d}N_{\text{ch}}/\textrm{d}\eta$ are measured in the forward region covered by the TOTEM T2 telescopes ($5.3 < |\eta| < 6.4$) for primary tracks 
with $p_T > 40$~MeV, and in the central region covered by the CMS tracker ($|\eta| < 2.2$) for primary tracks with $p_T > 100$ MeV. In order to limit the effects of pileup, events with more than one 
reconstructed vertex are rejected. The pseudorapidity distributions are measured for 3 different event topologies selected offline by the following requirements. A sample of inclusive inelastic events
is defined by the presence of at least one primary track candidate in one of the T2 telescopes in either side of the IP. A sample enhanced in non single diffractive dissociation (NSD) events is selected 
by requiring at least one primary track candidate in both T2 telescopes, and a sample enriched in single diffractive dissociation (SD) events is obtained by requiring at least one primary track candidate 
in one T2 telescope and none in the other. For the completeness of the discussion, the distributions associated to the 3 different topologies are presented. \\

The $\textrm{d}N_{\text{ch}}/\textrm{d}\eta$ distributions are corrected for the primary track reconstruction, the vertex reconstruction and selection, the event selection, the misidentification of secondary
tracks as primary ones and the multiple reconstruction of single charged particles. The final measurements are extrapolated down to $p_T = 0$ with correction factors determined in a MC-driven way from the 
$p_T$ spectrum of primary charged particles. For the measurement in the central region, the dominant sources of systematic uncertainties for the inclusive and NSD-enhanced samples are the track 
reconstruction and the event selection. The total systematic uncertainty ranges from 5 to 7 $\%$ and 6 to 8 $\%$ for the inclusive and NSD-enhanced samples, respectively. For the SD-enhanced sample, 
the dominant systematic uncertainties are due to the uncertainty in the event selection and the model dependence of the event selection efficiency. The total systematic uncertainty is between 10 and 
17 $\%$. For the measurement in the forward region, the dominant sources of systematic uncertainties for the inclusive and NSD-enhanced samples are the primary track selection, the track reconstruction 
and the description of the material between the T2 telescopes and the IP. The total systematic uncertainty ranges from 10 to 12 $\%$ for both samples. For the SD-enhanced sample, the most significant 
systematic uncertainty is due to the event selection, and the total systematic uncertainty is between 16 and 18 $\%$. \\

\begin{figure}[htb!]
\begin{center}
\includegraphics[width=0.45\textwidth, height=0.29\textheight]{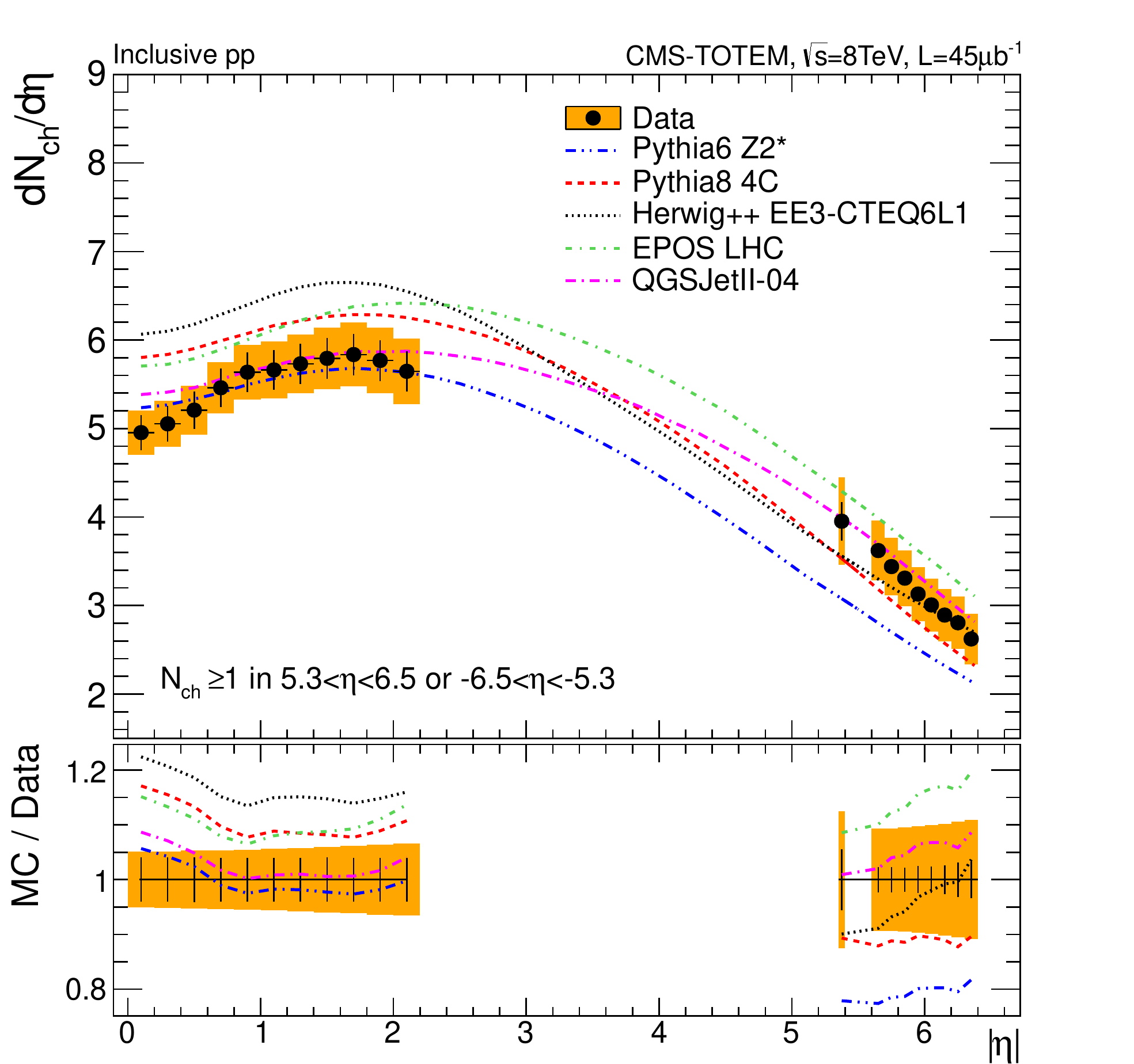}
\includegraphics[width=0.45\textwidth, height=0.29\textheight]{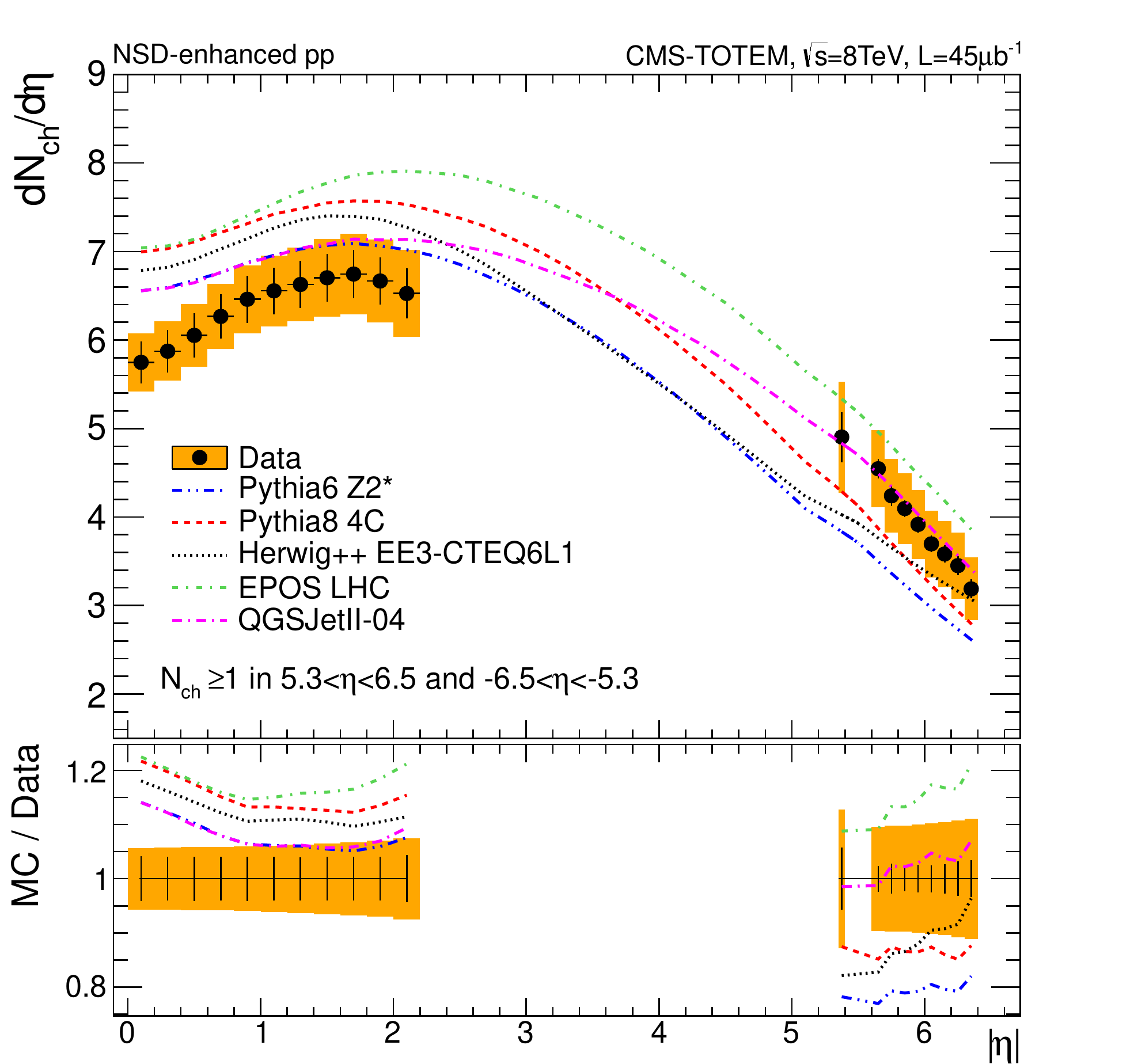}\\
\includegraphics[width=0.5\textwidth,height=0.29\textheight]{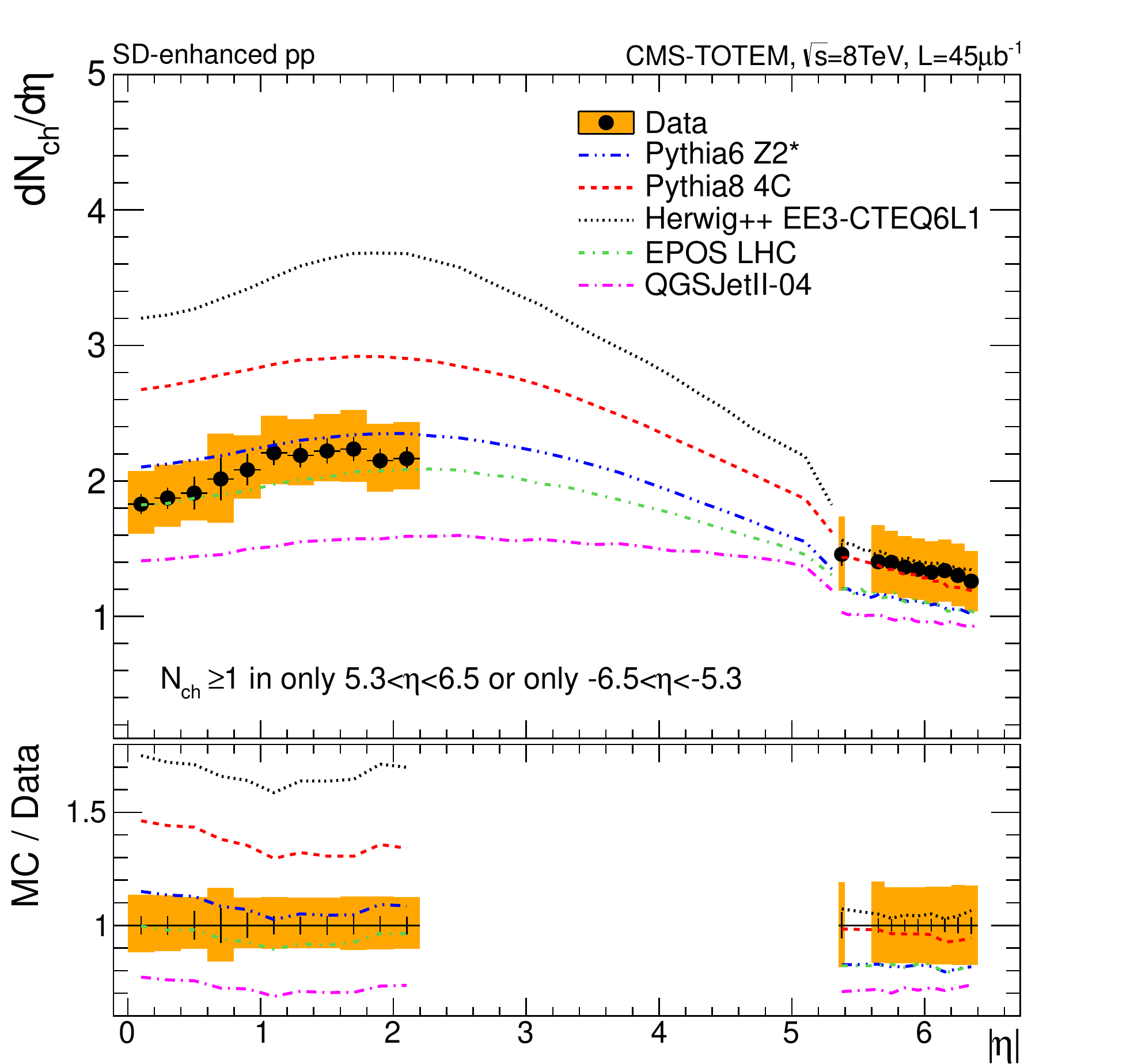}
\caption{Pseudorapidity distributions of charged particles for an inclusive sample (top left), a NSD-enhanced sample (top right), and a SD-enhanced sample (bottom). The error bars indicate 
the statistical and uncorrelated systematic uncertainties added in quadrature, while the bands show the total uncertainty. The data are compared to predictions of the MC event generators 
{\sc PYTHIA}6-Z2$^*$, {\sc PYTHIA}8-4C, {\sc HERWIG++} with tune UE-EE-3, {\sc EPOS} with tune LHC, and {\sc QGSJet}II-04.} 
\label{fig:CMSTOTEM}
\end{center}
\end{figure}

The combined CMS-TOTEM measurement of pseudorapidity distributions of charged particles is presented in figure~\ref{fig:CMSTOTEM}. The results are obtained by averaging the data points in the 
corresponding positive and negative $\eta$ bins. The band represents the total uncertainty, while the error bars indicate the statistical and uncorrelated systematic uncertainties added in quadrature.
The data are compared to predictions of the MC event generators {\sc PYTHIA}6-Z2$^*$, {\sc PYTHIA}8-4C and {\sc HERWIG++}~\cite{Bahr:2008pv} that has a recent tune to LHC data, 
UE-EE-3~\cite{Gieseke:2012ft}. The data are also compared to predictions of two MC event generators used in cosmic ray physics, {\sc EPOS}~\cite{Werner:2005jf} with the LHC tune~\cite{Pierog:2013ria} 
and {\sc QGSJet}II-04~\cite{Ostapchenko:2010vb}. These two MC models include contributions from soft and hard parton dynamics. The soft component is described in terms of the exchange of virtual 
quasi-particle states, while the hard component is included via hard-Pomeron scattering diagrams, which are equivalent to a leading-order perturbative QCD approach with DGLAP evolution. The models
used in {\sc EPOS} and {\sc QGSJet}II-04 are retuned 
 \newline
to LHC data~\cite{Pierog:2013jp}. None of the models considered provide a consistent description of the measured distributions.
In the central region, the predictions from the various MC event generators differ from the data by up to 20 $\%$ for the inclusive and NSD-enhanced samples, with even larger discrepancies for the 
SD-enhanced sample. For the inclusive sample, the data are well described by {\sc PYTHIA}6 and {\sc QGSJET}II-04 . For the NSD-enhanced sample, the predictions from {\sc PYTHIA}6 and {\sc QGSJET}II-04 
agree with the data for most $\eta$ bins. For the SD-enhanced sample, both {\sc EPOS} and {\sc PYTHIA}6 provide a good description of the measurement. In the forward region, the MC predictions differ 
from the data by up to 30 $\%$ for the three samples. For the inclusive and NSD-enhanced samples, the data are in agreement with the predictions from {\sc QGSJET}II-04, and between the {\sc EPOS} 
and {\sc PYTHIA}8 results. For the SD-enhanced sample, the TOTEM data points are close to the {\sc PYTHIA}8 and {\sc HERWIG++} predictions, while {\sc QGSJET}II-04 underestimates the data. 

\section{Prospects of common CMS-TOTEM data taking}
\label{sec:prospects}

During the LHC Run 1, diffractive processes have been mainly selected by requiring a large rapidity gap in the event. As previously mentioned, the successful collaboration between 
the CMS and TOTEM experiments has recently enabled us to complement the central CMS measurement with the scattered proton information, measured with the TOTEM Roman pots, in dedicated common runs 
during 2012 and 2013. Tagging of the scattered proton(s) enables to measure the four-momentum squared exchanged at the proton vertex, $t$, as well as the fractional momentum loss of the proton, $\xi$, 
when studying SD and CEP processes, and to reduce significantly the proton dissociation background that is present in measurements with the CMS central detector only. In the case of CEP, tagging 
the scattered protons allows reconstructing the mass of the centrally produced system with a resolution not achievable with the central detector only. \newline
 
A common CMS-TOTEM project for a precision proton spectrometer (CT-PPS)~\cite{CT-PPS} has now been approved, with installation planned at the end of the ongoing Long Shutdown 1 (LS1). This project
consists of tracking and timing detectors and is comprised of two phases, with the first taking place in 2015 at the start of the LHC Run 2, and the second expected to start one year later. Data will be
collected at high instantaneous luminosity with a standard low $\beta^*$ value of 0.5 m. The collected luminosity is expected to be of the order of 100 fb$^{-1}$ before the start of the Long Shutdown 2
(LS2) in 2018. \newline

In Phase I of the CT-PPS project, data will be collected by making use of the existing 147~m TOTEM horizontal Roman pots, relocated in the 200-225 m region and instrumented with silicon strip tracking
detectors. The tracking system will be complemented by timing detectors located in a new cylindrical Roman pot, which make it possible to identify the vertex associated to the scattered protons.
At high instantaneous luminosity, each crossing of the proton beams has an increased probability to produce several $pp$ collisions, and the availability of the timing detectors is critical in the
presence of a high pileup environment. The data acquisition of TOTEM is expected to be fully integrated with that of CMS, opening the possibility to combine CMS and TOTEM information at trigger level.
This first phase will be essentially dedicated to the understanding of the detector and backgrounds. The silicon strip detectors being not able to survive a continuous data taking at high instantaneous
luminosity, they will be mainly used to perform rate measurements during short insertions at the end of the fills. The integrated luminosity collected during the first phase of the CT-PPS project will
only correspond to a small fraction of the expected 100 fb$^{-1}$. \newline

In Phase II of the CT-PPS project, the silicon strip tracking detectors will be replaced by radiation hard pixel detectors. This second phase will be devoted to the study of small cross section high mass CEP
processes with the mass of the centrally produced system higher than 300 GeV. Central exclusive production provides a unique method to access a variety of physics topics, such as new physics via anomalous
production of $W$ and $Z$ boson pairs, high transverse momentum dijet production, and possibly the production of new resonances. \newline

Apart from the CT-PPS project, for which the hardware is jointly built, managed and operated by CMS and TOTEM together, the TOTEM collaboration will pursue in 2015 a high cross section forward physics
programme at low instantaneous luminosity~\cite{TOTEM-alone}. The collected luminosity is expected to be of the order of 1-10 pb$^{-1}$ for a high $\beta^*$ value of 90 m. These special runs will be
supported by CMS as common runs in terms of trigger and detector readout, in order to complement the central CMS measurement with the scattered proton information measured with the TOTEM Roman pots.
The physics programme is dedicated to the study of low mass SD and CEP processes, with the mass of the diffractive or centrally produced system smaller than 100 GeV. These processes include the single
diffractive production of a dijet system, $J/\Psi$ or $W$ boson, the CEP production of a dijet system, and the search for glueballs. 

\section{Summary}
\label{sec:summary}

A review of the CMS results on diffraction has been presented. These include the measurements of the soft diffractive cross sections~\cite{CMS:2013mda}, of the forward rapidity gap 
cross section~\cite{CMS:2013mda}, of the diffractive dijet cross section~\cite{Chatrchyan:2012vc}, the measurement of a large rapidity gap in $W$ and $Z$ boson events~\cite{Chatrchyan:2011wb}, 
and the measurement of the pseudorapidity distribution of charged particles in a single diffractive enhanced sample~\cite{Chatrchyan:2014qka}. This last measurement is the first common result of the CMS 
and TOTEM collaborations. Given the importance of the diffractive contribution to the UE and the need to constrain its phenomenological description, the measurement of the soft diffractive cross sections 
is of primary importance. The use of the forward CASTOR calorimeter has enabled us to measure both the SD and DD cross sections, which provide valuable input to constrain the modelling of hadronization 
and diffraction. The measurement of the forward rapidity gap cross section is consistent with the result of the ATLAS collaboration~\cite{Aad:2012pw} and extends the ATLAS measurement by 0.4 unit 
of gap size. The first observation of a diffractive contribution to dijet production at the LHC has enabled to obtain an estimate of the rapidity gap survival probability compatible with the CDF 
measurement~\cite{Affolder:2000vb,Aaltonen:2012tha}. The use of the LO MC event generators {\sc POMPYT} and {\sc POMWIG} gives a value of $0.12 \pm 0.05$, while the use of the NLO predictions based 
on {\sc POWHEG} gives a lower value of $0.08 \pm 0.04$. The measurement of a large rapidity gap in $W$ and $Z$ boson events is the first observation of a diffractive contribution to $W$ and $Z$ bosons
production at the LHC. The relative diffractive contribution in a sample of W boson events with a LRG is found to be $(50.0 \pm 9.3 \, (\mbox{stat.}) \pm 5.2 \, (\mbox{syst.})) \%$ when using a combination
of the MC event generators {\sc PYTHIA}6 Pro-Q20 and {\sc POMPYT} to describe the nondiffractive and single diffractive components, respectively. The measurement of pseudorapidity distributions of charged
particles covers the largest pseudorapidity interval ever measured at the LHC and has the potential to probe the correlation between particle production in the central and forward regions. None of the
models considered provide a consistent description of the measurement over the whole $\eta$ range and for the 3 different event samples.

\section{Acknowledgements}

The author would like to thank the members of the diffractive and exclusive group of the CMS collaboration, as well as the members of the TOTEM collaboration involved in the common CMS-TOTEM analyses. 
A special thank to Pierre Van Mechelen, Hans Van Haevermaet, Robert Ciesielski and Kenneth {\"O}sterberg for the fruitful discussions on the ongoing analyses.

\end{document}